\documentclass[12pt,draftclsnofoot,onecolumn]{IEEEtran}
\usepackage[T1]{fontenc}
\usepackage[latin9]{inputenc}
\usepackage{xcolor}
\usepackage{array}
\usepackage{float}
\usepackage{units}
\usepackage{multirow}
\usepackage{amsmath}
\usepackage{amssymb}
\usepackage{graphicx}
\usepackage{setspace}
\usepackage[unicode=true,
 bookmarks=true,bookmarksnumbered=true,bookmarksopen=true,bookmarksopenlevel=1,
 breaklinks=false,pdfborder={0 0 0},pdfborderstyle={},backref=false,colorlinks=false]
 {hyperref}
\hypersetup{pdftitle={Your Title},
 pdfauthor={Your Name},
 pdfpagelayout=OneColumn, pdfnewwindow=true, pdfstartview=XYZ, plainpages=false}

\makeatletter

\providecommand{\tabularnewline}{\\}
\floatstyle{ruled}
\newfloat{algorithm}{tbp}{loa}
\providecommand{\algorithmname}{Algorithm}
\floatname{algorithm}{\protect\algorithmname}

\usepackage[caption=false,font=footnotesize]{subfig}
\usepackage{stfloats}
\usepackage{algorithm}
\usepackage{algorithmic}
\usepackage{amsmath}

\@ifundefined{showcaptionsetup}{}{%
 \PassOptionsToPackage{caption=false}{subfig}}
\usepackage{subfig}
\makeatother

\begin{document}

\title{Design and Performance Analysis of Multi-scale NOMA for 5G Positioning}

\author{Lu~Yin,~\IEEEmembership{Member,~IEEE,}~Jiameng~Cao,~Zhongliang~Deng,~Qiang~Ni,~\IEEEmembership{Senior Member,~IEEE},~Song~Li,~\IEEEmembership{Member,~IEEE,}~Xinyu~Zheng,~Hanhua~Wang\thanks{Lu~Yin,~Jiameng~Cao,~Zhongliang~Deng,~Xinyu~Zheng, and Hanhua~Wang
are with the School of Electronic Engineering, Beijing University
of Posts and Telecommunications, Beijing, 100876, China, e-mail: \protect\href{http://inlu_mail@bupt.edu.cn}{inlu\_mail@bupt.edu.cn},
\protect\href{http://caojiameng@bupt.edu.cn}{caojiameng@bupt.edu.cn},
\protect\href{http://dengzhl@bupt.edu.cn}{dengzhl@bupt.edu.cn}, \protect\href{http://buptzxy@bupt.edu.cn}{buptzxy@bupt.edu.cn},
\protect\href{http://whh0710@bupt.edu.cn}{whh0710@bupt.edu.cn}.}\thanks{Qiang~Ni is with the School of Computing and Communications, Lancaster
University, InfoLab21, LA1 4WA, U.K. e-mail: \protect\href{http://q.ni@lancaster.ac.uk}{q.ni@lancaster.ac.uk}.}\thanks{Song~Li is with School of Information and Control Engineering, China
University of Mining and Technology, Xuzhou, China. e-mail: \protect\href{http://lisong@cumt.edu.cn}{lisong@cumt.edu.cn}.}}
\maketitle
\begin{abstract}
This paper presents a feasibility study for a novel positioning-communication
integrated signal called Multi-Scale Non-Orthogonal Multiple Access
(MS-NOMA) for 5G positioning. One of the main differences between
the MS-NOMA and the traditional positioning signal is MS-NOMA supports
configurable powers for different positioning users (P-Users) to obtain
better ranging accuracy and signal coverage. Our major contributions
are: Firstly, we present the MS-NOMA signal and analyze the Bit Error
Rate (BER) and ranging accuracy by deriving their simple expressions.
The results show the interaction between the communication and positioning
signals is rather limited, and it is feasible to use the MS-NOMA signal
to achieve high positioning accuracy. Secondly, for an optimal positioning
accuracy and signal coverage, we \textcolor{black}{model the power
allocation problem for MS-NOMA signal as a convex optimization problem}
by satisfying the QoS (Quality of Services) requirement and other
constraints. \textcolor{black}{Then, we propose a novel} Positioning-Communication
Joint Power Allocation (PCJPA) algorithm\textcolor{black}{{} which allocates
the powers of all P-Users iteratively.} The theoretical and numerical
results show our proposed MS-NOMA signal has great improvements of
ranging/positioning accuracy than traditional PRS (Positioning Reference
Signal) in 5G, and improves the coverage dramatically which means
more P-Users could locate their positions without suffering the near-far
effect.
\end{abstract}


\section{Introduction}

Nowadays, the Location Based Services (LBS) are growing rapidly and
attracting much attention with the proliferation of mobile devices
\cite{del2018Survey,Shojafar2016Energy}. The well known Global Navigation
Satellite Systems (GNSS), such as the Global Positioning System (GPS)
and the Beidou System (BDS) \cite{Liang2012Automated,7279407}, can
only be used in open areas\textcolor{black}{{} as their signals are
}easily blocked or interfered by buildings\textcolor{black}{{} \cite{Yin2018A}.
The emerging Wi-Fi, Bluetooth or Wireless Sensor Networks (WSN) based
positioning have well coverage indoors only with dense placements
of the nodes \cite{Jeon2016An,Tomic2014RSS,Yin2019Intelligent}. And
it is costly for collecting and maintaining the fingerprint database}
as well \cite{ZouWinIPS,Shuai2016Automatic}.

\textcolor{black}{Wireless communication network has a well coverage
both indoors and outdoors \cite{Chen2018Coverage,Zhou2018Coverage}.
Meanwhile, it is cost-effective to be used for positioning purpose
as it is ready-made for communication purpose. }However, the positioning
accuracy can not meet some high-accuracy requirements by using the
communication signal directly as the communication system is not designed
for positioning purpose specifically \cite{Vaghefi2016Cooperative,Cui2017Real}.
For example, the Positioning Reference Signal (PRS) in the cellular
network cannot fully meet the commercial requirements as the discontinuous
signal is hardly tracked which leads to a low range measurement accuracy
\cite{3GPPTR38855}. \textcolor{black}{Moreover, there are severe
near-far effects between the positioning signals from different gNBs
(next generation NodeBs or called 5G nodeBs) which makes the signals
from far gNBs more difficult to be received \cite{Schloemann2015A,Deng2013Situation,Manna2016Performance}.
Consequently, poor geometric distribution of gNBs is achieved which
further worsen the positioning accuracy \cite{Schloemann2015AT,Chen2015Enhanced}.}

\begin{figure}[tbh]
\begin{centering}
\includegraphics[viewport=20bp 20bp 280bp 190bp,clip,width=0.5\columnwidth]{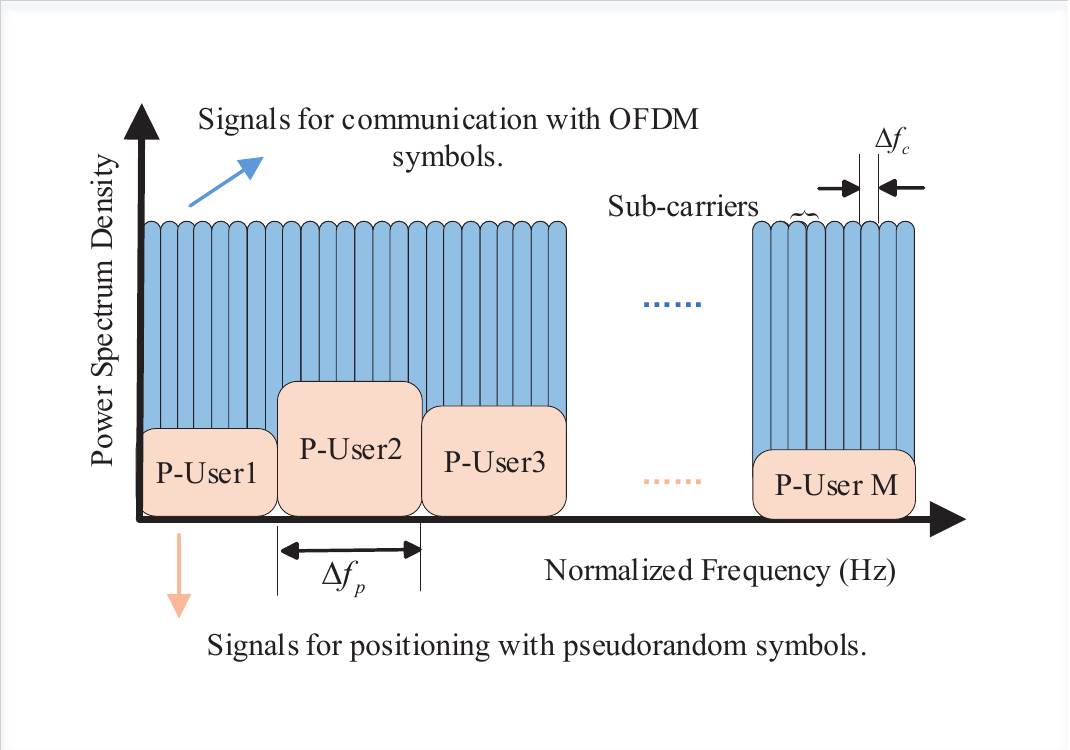}
\par\end{centering}
\caption{The MS-NOMA \textcolor{black}{architecture in frequency and power
domains\label{fig:The-proposed-FD-NOMA-1}}}
\end{figure}

To this end, we propose a positioning-communication integrated signal,
called Multi-Scale Non-Orthogonal Multiple Access (MS-NOMA), \textcolor{black}{which
superposes a low power positioning signal to the communication one
without much interference} based on the NOMA principle. In time domain,
the MS-NOMA signal is Code Division Multiple Access (CDMA) to obtain
corresponding spreading gains and ensure a continuous transmission.
In frequency domain, Orthogonal Frequency Division Multiplexing Access
(OFDMA) is employed for different positioning users as Fig. \ref{fig:The-proposed-FD-NOMA-1}
shows.

\textcolor{black}{In the proposed MS-NOMA signal, $\varDelta f_{c}$
and $\varDelta f_{p}$ represent the sub-carrier spacing of }communication
user (C-User)\textcolor{black}{{} and }positioning user (P-User)\textcolor{black}{,
respectively. For the maximum spectrum effectiveness, they are designed
as $\varDelta f_{p}=G\varDelta f_{c},\thinspace G\in\mathbb{N}_{+}$.
We assume both C-User and P-User each occupy a separate sub-carrier,
so that there are maximum $N=B/\varDelta f_{c}-1$ and $M=B/\varDelta f_{p}-1$
users for communication and positioning purpose, respectively. Where
$B$ represents the total bandwidth.} The reasons for distinguishing
different P-Users are:
\begin{enumerate}
\item Unlike communication system, it usually needs more than three gNBs
to calculate the P-User's position. As the P-Users are located at
different locations, the powers of the positioning signals from far
gNBs must be high enough in order to be correctly recognized. Meanwhile,
the powers of the positioning signals from near gNBs should be low
enough to avoid the near-far effect. So different powers for different
P-Users are necessary.
\item The superposed positioning signal interferes the communication signal
like inter-user interference occurred in the normal NOMA as well \cite{Yu2018Link,Liu2017Nonorthogonal}.
To reduce this interference, the power of the positioning signal must
be limited under a certain threshold to satisfy the QoS (Quality of
Services) requirement of communication. While for P-Users, higher
powers are needed for more accurate range measuring \cite{Yin2019Anovel}.
Therefore, the gNBs could transmit positioning signals with different
powers for different P-Users to meet the requirements of both C-Users
and P-Users.
\end{enumerate}
Although the aforementioned problems may be mitigated by varying the
bandwidths of the positioning signals as well, it will bring some
other problems and make the whole system more complicated. In this
paper, we discuss the scenario that the bandwidths of all P-Users
are identical and fixed. Then, to acquire the highest positioning
accuracy over the whole network with the hearability and QoS requirements,
the powers of different P-Users must be allocated care\textcolor{black}{fully.}

\textcolor{black}{In}\textcolor{brown}{{} }\textcolor{black}{a conventional
OFDM system, it is proved that water-filling over the sub-carriers
is the optimal power allocation strategy \cite{Karmokar2015Energy,6758420}.
However, it does not consider the interferences between different
types of users. In \cite{Ding2017Joint}, where the second user transmit
over spectrum holes left in the primary system, an optimal power allocation
strategy is proposed. They maximize the down-link capacity of the
second user by remaining the interference introduced to the primary
user within a tolerable range. In the NOMA system, the power allocation
is mostly investigated for signal demodulation and relay transmission}\textcolor{brown}{{}
}\textcolor{black}{\cite{Yan2018Toward,7954630}. But these algorithms
can not be used in our problem that has different models which is
more complicated.}

To the best of our knowledge, there are few studies about the power
allocation in positioning \textcolor{black}{systems}\textcolor{blue}{{}
}which makes our study very challenging. A preliminary part of this
study was presented in a letter paper \cite{Yin2019Anovel}. In this
journal version, as compared to \cite{Yin2019Anovel}, we carry out
detailed design and conduct more in-depth mathematical performance
analysis. The main contributions of this paper are:
\begin{enumerate}
\item We present the system model of positioning by the proposed MS-NOMA
signal and analyze the limitation of positioning by other existing
signals.
\item We analyze the interferences between the communication and positioning
signals of the MS-NOMA signal in the multi-cell scenario. Bit Error
Rate (BER) for communication and the ranging error for positioning
are derived.
\item \textcolor{black}{We model the power allocation problem for the MS-NOMA
signal as a convex optimization problem. It minimizes the average
positioning error of all P-Users in the network by considering the
QoS, the power budget and the }hearability\textcolor{black}{{} }requirements\textcolor{black}{.
To solve this problem, we propose a novel} Positioning-Communication
Joint Power Allocation (PCJPA) algorithm\textcolor{black}{{} which allocates
the powers of all P-Users iteratively and derive its solution.}
\item A series of theoretical and numerical analysis are done to evaluate
the feasibility of positioning by MS-NOMA signal. The results show
our proposed MS-NOMA signal has a great improvement of ranging accuracy
than PRS in traditional 5G signal, and improves the coverage dramatically
which means more P-Users could locate their positions without suffering
the near-far effect.
\end{enumerate}
\textcolor{black}{Notations: $\left\Vert \cdot\right\Vert $ represents
the Euclidean distance. The operator $\textrm{cov}\left(\cdot\right)$
represents the covariance. $kn$ and $km$ represent the $n^{\textrm{th}}$
C-User/communication signal and the $m^{\textrm{th}}$ P-User/positioning
signal served/broadcast by gNB $k$, respectively. $\mathcal{M}$,
$\mathcal{N}$, $\mathcal{K}$ and $\mathcal{K}^{k}$ represent the
set $\left\{ 1,...,M\right\} $, $\left\{ 1,...,N\right\} $, $\left\{ 1,...,K\right\} $
and $\left\{ 1,...,k-1,k+1,...,K\right\} ,$respectively.}

\section{System Model}

\begin{figure}[tbh]
\begin{centering}
\includegraphics[viewport=20bp 20bp 530bp 420bp,clip,width=0.7\columnwidth]{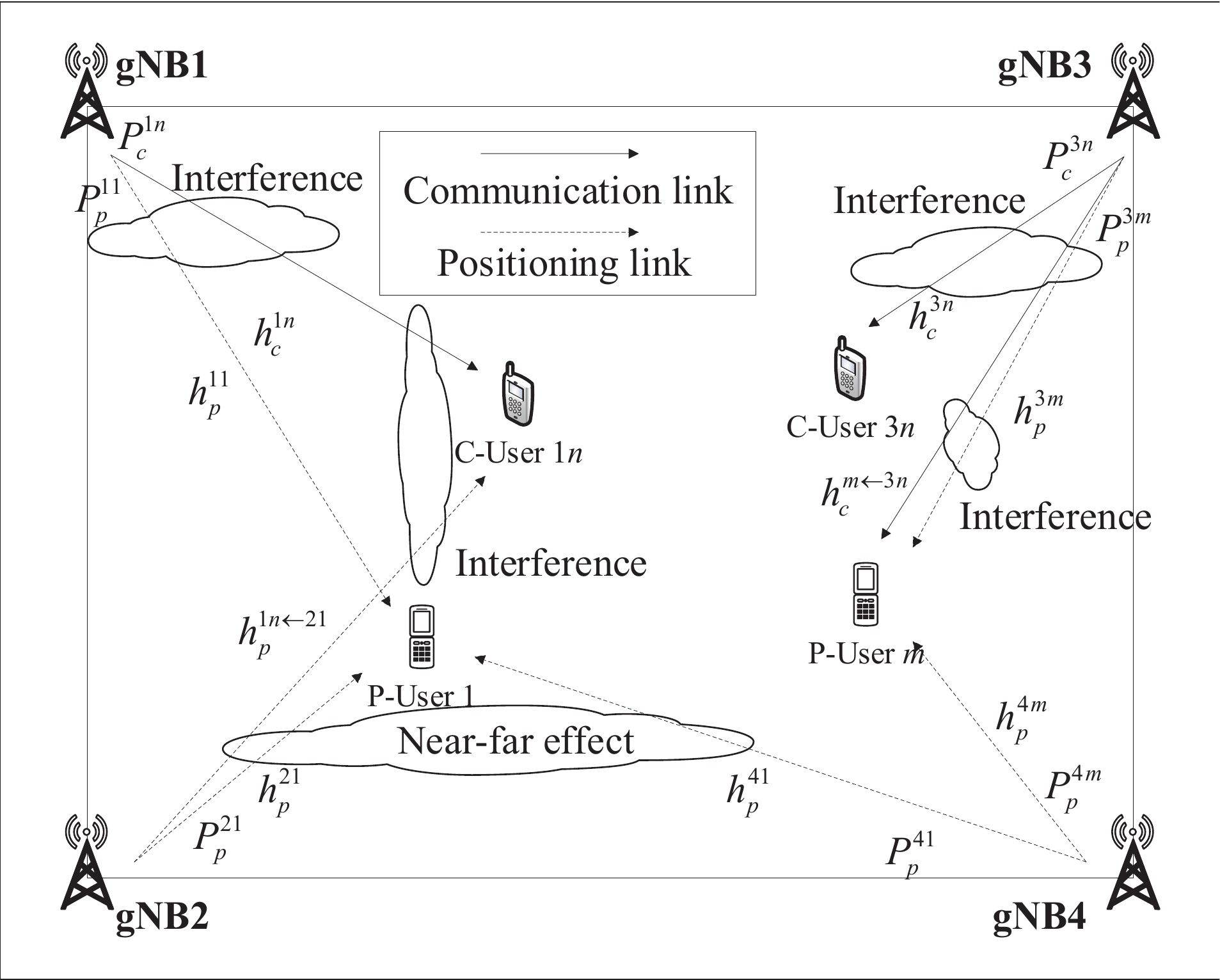}
\par\end{centering}
\caption{System model of positioning by the MS-NOMA signal\label{fig:Illustration-of-localization}}
\end{figure}

Consider a typical positioning scenario in spatial domain with $K$
gNBs as Fig. \ref{fig:Illustration-of-localization} shows. C-Users
and P-Users may be located at different locations. Because each C-User
connects only one gNB, there are maximum $KN$ C-Users in the network.
While a P-User needs as many connections as possible for accurate
positioning. So, there are maximum $M$ P-Users in the network under
the assumption that all gNBs server the same P-Users.

To ensure the P-Users receiving more than one gNB's signal, the powers
of the positioning signals should be strong enough. Then, this strong
positioning signal may interfere the neighbor C-Users. To examine
these interferences, let us define $h_{c}^{kn}$ and $h_{p}^{km}$
as the instantaneous channel gains between gNB $k$ and C-User $kn$/P-User
$m$, respectively. N\textcolor{black}{otice that the communication/positioning
signal broadcast from gNB $k'$ to the C-User $k'n$/P-User $m$ will
received by the P-User $m$/C-User $kn$ as well, we use $h_{c}^{m\leftarrow k'n}$
and $h_{p}^{kn\leftarrow k'm}$ to represent these two instantaneous
channel gains, respectively.}

\textcolor{black}{Without any loss of generality, we assume: }1) Each
spreading sequence for different P-Users is independent;\textcolor{black}{{}
2) The powers for all C-Users are identical and }the powers \textcolor{black}{for
all P-Users} are to be allocated\textcolor{black}{; and 3)} The channel
states are available through a delay- and error-free feedback channel
which are known by the gNBs\textcolor{black}{.}

What we concern is the horizontal positioning accuracy\footnote{The vertical positioning accuracy of terrestrial system is usually
very larger as the large vertical dilution \cite{Zhong2009Geometric}.
Other methods are usually used to estimate vertical position rather
than terrestrial positioning system \cite{Jeon2016An,Deng2013Situation}.}. If the gNBs are perfectly synchronized, the P-Users will use time-based
algorithm to fix their locations. Then, the horizontal positioning
accuracy of P-User $m$ can be expressed as
\begin{equation}
\varPsi^{m}=\sqrt{\sum_{k\in\mathcal{K}}\left(\lambda^{km}\sigma_{\rho}^{km}\right)^{2}}\label{eq:1-2}
\end{equation}
where $\lambda^{km}$ and $\sigma_{\rho}^{km}$ are given in Appendix
\ref{subsec:Derivation-of-the}.

\section{Features of MS-NOMA Signal}

\subsection{Interference of the Positioning Signal to Communication One}

\textcolor{black}{For evaluating the }interference of the positioning
signal to communication one, we assume the inter-cell interference
between the communication signals could be ideally eliminated. Then,
the BER of C-User $kn$ is \cite{Liu2016Cooperative}

\textcolor{black}{
\begin{equation}
BER^{kn}=\Gamma\textrm{erfc}\left(\frac{\gamma\left|h_{c}^{kn}\right|^{2}P_{c}T_{c}}{I^{kn}+2N_{0}}\right)\label{eq:1}
\end{equation}
where $\Gamma$ and $\gamma$ are determined by the modulation and
coding schemes. $P_{c}$ is the power of the communication signal.
$T_{c}$ is the period of the communication symbol. $N_{0}$ is the
environment noise\textquoteright s single-sided Power Spectral Density
(PSD). $I^{kn}$ represents the interference of the positioning signal
to the C-User $kn$. }Notice that the powers of the positioning signals
from different gNBs may be similar by power allocation, the interferences
caused by the positioning signals from the neighbor gNBs can not be
ignored\textcolor{black}{. Then we have
\begin{equation}
I^{kn}=\sum_{k'\in\mathcal{K}}\sum_{m\in\mathcal{M}}\bar{P}_{p}^{kn\leftarrow k'm}\label{eq:2}
\end{equation}
where $\bar{P}_{p}^{kn\leftarrow k'm}$ is the power of the positioning
signal $k'm$ received by C-User $kn$ which satisfies}

\textcolor{black}{
\begin{align}
\bar{P}_{p}^{kn\leftarrow k'm} & =\left|h_{p}^{kn\leftarrow k'm}\right|^{2}P_{p}^{k'm}G_{p}^{m}\left(n\varDelta f_{c}\right)\nonumber \\
 & =\left|h_{p}^{kn\leftarrow k'm}\right|^{2}P_{p}^{k'm}T_{p}\textrm{sinc}^{2}\left(m-\frac{n}{G}\right)\label{eq:4}
\end{align}
where $P_{p}^{k'm}$ is the power of positioning signal $m$ broadcast
by gNB $k'$. $G_{p}^{m}\left(f\right)$ is the normalized PSD of
positioning signal $m$ which satisfies
\begin{equation}
G_{p}^{m}\left(f\right)=T_{p}\textrm{sinc}^{2}\left[\left(f-m\Delta f_{p}\right)T_{p}\right]\label{eq:5-1}
\end{equation}
where $T_{p}$ is the period of the positioning symbol.}

\subsection{Ranging Accuracy of MS-NOMA Signal}

The receiver could use a Delay Locked Loop (DLL) to track the positioning
signal. Taking the coherent early-late discriminator for example \cite{Betz2009Generalized},
the tracking/ranging error of the positioning signal $km$ can be
written as
\begin{multline}
\left(\sigma_{\rho}^{km}\right)^{2}=\\
\frac{a\intop_{B_{0}-B_{fe}/2}^{B_{0}+B_{fe}/2}\left[N_{0}+G_{s}^{m}\left(f+m\Delta f_{p}\right)+G_{q}^{km}\left(f+m\Delta f_{p}\right)\right]G_{p}^{m}\left(f+m\Delta f_{p}\right)\textrm{sin}^{2}\left(\pi fDT_{p}\right)df}{\left|h_{p}^{km}\right|^{2}P_{p}^{km}\left[2\pi\intop_{B_{0}-B_{fe}/2}^{B_{0}+B_{fe}/2}fG_{p}^{m}\left(f+m\Delta f_{p}\right)\textrm{sin}\left(\pi fDT_{p}\right)df\right]^{2}}\label{eq:3}
\end{multline}
 where\textcolor{black}{{} }$a$ is determined by the loop parameters.
$B_{0}$ is the central frequency of MS-NOMA signal. $B_{fe}$ is
the double-sided front-end bandwidth. $D$ is the early-late spacing
of DLL. $G_{s}^{m}\left(f\right)$ is the PSD of the communication
signals received by P-User $m$ which satisfies
\begin{align}
G_{s}^{m}\left(f\right) & =\sum_{k'\in\mathcal{K}}\sum_{n\in\mathcal{N}}\left|h_{c}^{m\leftarrow k'n}\right|^{2}P_{c}G_{c}^{n}\left(f\right)\label{eq:5}
\end{align}
\textcolor{black}{where $G_{c}^{n}\left(f\right)=T_{c}\textrm{sinc}^{2}\left[\left(f-n\Delta f_{c}\right)T_{c}\right]$
}is the \textcolor{black}{normalized PSD of communication signal $n$.
$G_{q}^{km}\left(f\right)$ is the PSD of the positioning signals
from other gNBs} which satisfies
\begin{equation}
G_{q}^{km}\left(f\right)=\sum_{k'\in\mathcal{K}^{k}}\left|h_{p}^{k'm}\right|^{2}P_{p}^{k'm}G_{p}^{m}\left(f\right)\label{eq:8}
\end{equation}

\textcolor{black}{By taking (\ref{eq:5-1}), (\ref{eq:5})-(\ref{eq:8})
into (\ref{eq:3}) and using some approximations, (\ref{eq:3}) can
be simplified to
\begin{align}
\left(\sigma_{\rho}^{km}\right)^{2} & \approx\frac{aT_{p}^{2}}{2}\left[\frac{1}{B_{fe}T_{p}\left(C/N_{0}\right)^{km}}+\frac{B\sum_{k'\in\mathcal{K}}\left(CPR\right)^{km\leftarrow k'}}{2B_{fe}^{2}}+\frac{\sum_{k'\in\mathcal{K}^{k}}\left(PPR\right)^{km\leftarrow k'm}}{B_{fe}^{2}T_{p}}\right]\label{eq:7-1}
\end{align}
where $\left(C/N_{0}\right)^{km}$, }$\left(CPR\right)^{km\leftarrow k'}$
and $\left(PPR\right)^{km\leftarrow k'm}$ can be found in \textcolor{black}{Appendix
\ref{subsec:Derivation-of}}. \textcolor{black}{The first item in
(\ref{eq:7-1}) is caused by the noise, the second one is caused by
the communication signals from all gNBs, and the third one is caused
by the other gNBs' positioning signals.}

\textcolor{black}{We define the ranging-factor $\left(\widetilde{\sigma}_{\rho}^{km}\right)^{2}=\left(\sigma_{\rho}^{km}\right)^{2}P_{p}^{km}$
as (\ref{eq:9}) shows for later use.
\begin{align}
\left(\widetilde{\sigma}_{\rho}^{km}\right)^{2} & =\frac{aT_{p}^{2}}{2}\left[\frac{N_{0}}{B_{fe}T_{p}\left|h_{p}^{km}\right|^{2}}+\frac{BGP_{c}}{B_{fe}^{2}}\frac{\sum_{k'\in\mathcal{K}}\left|h_{c}^{m\leftarrow k'}\right|^{2}}{\left|h_{p}^{km}\right|^{2}}+\frac{\sum_{k'\in\mathcal{K}^{k}}\left|h_{p}^{k'm}\right|^{2}P_{p}^{k'm}}{B_{fe}^{2}T_{p}\left|h_{p}^{km}\right|^{2}}\right]\label{eq:9}
\end{align}
} 

\textcolor{black}{}

\section{The Power Allocation of MS-NOMA Signal\label{sec:Sum-range-measurement}}

\subsection{The Constraints}

\subsubsection{The BER Threshold under QoS Constraint}

To ensure the QoS of the C-Users, the BER of all C-Users should be
limited under a certain threshold
\begin{equation}
BER^{kn}\leq\Xi_{th},\thinspace\thinspace\thinspace\forall k\in\mathcal{K},\forall n\in\mathcal{N}\label{eq:6}
\end{equation}
Then, by taking (\ref{eq:1}) to (\ref{eq:6}) and rearranging items,
we have
\begin{align}
I^{kn} & \leq\frac{\gamma\left|h_{c}^{kn}\right|^{2}P_{c}T_{c}}{\textrm{erfc}^{-1}\left(\Xi_{th}/\Gamma\right)}-2N_{0}\nonumber \\
 & \triangleq I_{th}^{kn},\thinspace\thinspace\thinspace\forall k\in\mathcal{K},\forall n\in\mathcal{N}\label{eq:sub_1}
\end{align}
where $I_{th}^{kn}$ is defined as the interference threshold of C-User
$kn$ which is determined by the QoS requirement $\Xi_{th}$.

\subsubsection{The Total Power Limitation}

The total transmit power is often limited. In MS-NOMA signal, we have
\begin{equation}
\sum_{m\in\mathcal{M}}P_{p}^{km}+NP_{c}\leq P_{\textrm{T}}^{k},\thinspace\thinspace\forall k\in\mathcal{K}\label{eq:p}
\end{equation}
where $P_{\textrm{T}}^{k}$ is the total transmit power of gNB $k$.
Let's define\textcolor{black}{{} the positioning power budget of} gNB\textcolor{black}{{}
$k$ as} $P_{th}^{k}=P_{\textrm{T}}^{k}-NP_{c}$, then we have
\begin{equation}
{\displaystyle {\displaystyle \sum_{m\in\mathcal{M}}P_{p}^{km}\leq P_{th}^{k}}},\thinspace\thinspace\forall k\in\mathcal{K}\label{eq:sub_2}
\end{equation}

\subsubsection{The Elimination of Near-far Effect}

\textcolor{black}{To guarantee P-Users receive as many positioning
signals as possible, the powers of the received positioning signals
from different gNBs must satisfy
\begin{equation}
\frac{|h_{p}^{km}|^{2}P_{p}^{km}}{|h_{p}^{k'm}|^{2}P_{p}^{k'm}}\geq\varrho\Omega,\thinspace\thinspace\thinspace\forall m\in\mathcal{M},\forall k\in\mathcal{K},\forall k'\in\mathcal{K}^{k}\label{eq:sub_3}
\end{equation}
where $\Omega$ is the positioning signal's auto-correlation to cross-correlation
ratio which is determined by the pseudorandom code and its length.
$\varrho$ is determined by the receiver's performance which is usually
larger than 1. For a particular positioning signal $km$, if the strongest
cross-correlation satisfies (\ref{eq:sub_3}), all $k'$s in (\ref{eq:sub_3})
will be satisfied. So (\ref{eq:sub_3}) can be rewritten as
\begin{equation}
|h_{p}^{km}|^{2}P_{p}^{km}\geq\varrho\Omega|h_{p}^{k_{k}^{\prime}m}|^{2}P_{p}^{k_{k}^{\prime}m},\thinspace\thinspace\forall m\in\mathcal{M},\forall k\in\mathcal{K}
\end{equation}
where $k_{k}^{\prime}m$ represents the index of the strongest signal
received by the P-User $km$ except the positioning signal $km$.}

\subsection{The Proposed Joint Power Allocation Model}

\textcolor{black}{Our goal is to obtain a best positioning performance
for all P-Users in terms of both coverage and accuracy under QoS requirement
and total transmit power budget. So, the average }horizontal positioning
accuracy\footnote{For calculation convenience, we use the square of the horizontal positioning
accuracy, i.e. $\left(\varPsi^{m}\right)^{2}$}\textcolor{black}{{} for all P-Users in the network is minimized by
finding the power values $P_{p}^{km},\forall m\in\mathcal{M},\forall k\in\mathcal{K}$
under the given constraints.} We use the fact that maximization of
negative value of a convex function is equivalent to its minimization.
Mathematically, the power allocation problem can be formulated as
a convex optimization problem as follows

\begin{align}
\mathrm{OP1}: & \underset{P_{p}^{km}}{\max}-\frac{1}{M}\sum_{m\in\mathcal{M}}\left(\varPsi^{m}\right)^{2}\\
\textrm{s.t.} & I^{kn}\leq I_{th}^{kn},\thinspace\thinspace\forall n\in\mathcal{N},\forall k\in\mathcal{K}\\
 & \sum_{m\in\mathcal{M}}P_{p}^{km}\leq P_{th}^{k},\,\,\forall k\in\mathcal{K}\\
 & |h_{p}^{km}|^{2}P_{p}^{km}\geq\varrho\Omega|h_{p}^{k_{k}^{\prime}m}|^{2}P_{p}^{k_{k}^{\prime}m},\thinspace\thinspace\forall m\in\mathcal{M},\forall k\in\mathcal{K}\label{eq:19}
\end{align}

OP1 can be solved by the Lagrange duality method \cite{Boyd2006Convex}.
Then the Lagrangian of OP1 can be written as
\begin{align}
\mathcal{L} & \left(\left\{ P_{p}^{km}\right\} ,\mu,\nu,\beta\right)=-\frac{1}{M}\sum_{m\in\mathcal{M}}\sum_{k\in\mathcal{K}}\left(\lambda^{km}\sigma_{\rho}^{km}\right)^{2}+{\displaystyle \sum_{k\in\mathcal{K}}\sum_{n\in\mathcal{N}}\mu^{kn}\left(I_{th}^{kn}-I^{kn}\right)}\nonumber \\
 & \thinspace\thinspace\thinspace\thinspace\thinspace\thinspace\thinspace\thinspace\thinspace\thinspace\thinspace\thinspace\thinspace\thinspace\thinspace+\sum_{k\in\mathcal{K}}\nu^{k}\left(P_{th}^{k}-{\displaystyle \sum_{m\in\mathcal{M}}P_{p}^{km}}\right)+\sum_{m\in\mathcal{M}}\sum_{k\in\mathcal{K}}\beta^{km}\left(|h_{p}^{km}|^{2}P_{p}^{km}-\varrho\Omega|h_{p}^{k_{k}^{\prime}m}|^{2}P_{p}^{k_{k}^{\prime}m}\right)\label{eq:21}
\end{align}
where $\mu$, $\nu$ and $\beta$ are the matrices of dual variables
associated with the corresponding constraints given in (\ref{eq:sub_1}),
(\ref{eq:sub_2}) and (\ref{eq:sub_3})
\begin{equation}
\mu=\left\{ \mu^{kn},\forall k\in\mathcal{K},\forall n\in\mathcal{N}\right\} \in\mathbb{\mathbb{C}}^{K\times N}\succeq0
\end{equation}
\begin{equation}
\nu=\left\{ \nu^{k},\forall k\in\mathcal{K}\right\} \in\mathbb{\mathbb{C}}^{1\times K}\succeq0
\end{equation}
\begin{equation}
\beta=\left\{ \beta^{km},\forall k\in\mathcal{K},\forall m\in\mathcal{M}\right\} \in\mathbb{\mathbb{C}}^{K\times M}\succeq0
\end{equation}

The Lagrange dual function of OP1 is then given by

\begin{equation}
g\left(\mu,\nu,\beta\right)=\underset{P_{p}^{km}}{\max}\mathcal{L}\left(\left\{ P_{p}^{km}\right\} ,\mu,\nu,\beta\right)\label{eq:12}
\end{equation}
The dual optimization problem can be formulated as
\begin{align}
\mathrm{min} & g\left(\mu,\nu,\beta\right)\\
\textrm{s.t.} & \mu\succeq0,\thinspace\nu\succeq0,\thinspace\beta\succeq0
\end{align}
Obviously, $\mathcal{L}\left(\left\{ P_{p}^{km}\right\} ,\mu,\nu,\beta\right)$
is linear in $\mu,\nu,\beta$ for fixed $P_{p}^{km}$, and $g\left(\mu,\nu,\beta\right)$
is the maximum of linear functions. Thus, the dual optimization problem
is always convex. In the following, the dual decomposition method
introduced in \cite{Zhang2012Optimal} is employed to solve this problem.
For this purpose, we introduce a transformation $\sum_{n\in\mathcal{N}}=\sum_{m\in\mathcal{M}}\sum_{n\in\mathbb{N}_{m}}$
to decompose the Lagrange dual function to $K\times M$ independent
sub-problems, where 
\begin{equation}
\mathbb{N}_{m}=\left\{ \left(2G-1\right)(m-1)+1,...,\left(2G-1\right)m\right\} 
\end{equation}
Then we have

\begin{align}
g\left(\mu,\nu,\beta\right) & ={\displaystyle \sum_{k\in\mathcal{K}}\left[g^{k}\left(\mu,\nu,\beta\right)\right]}\label{eq:Fen_1}\\
 & ={\displaystyle \sum_{k\in\mathcal{K}}\left\{ \sum_{m\in\mathcal{M}}g^{km}\left(\mu,\nu,\beta\right)+\nu^{k}P_{th}^{k}\right\} }
\end{align}
where
\begin{align}
g^{km}\left(\mu,\nu,\beta\right) & =\underset{P_{p}^{km}}{\max}\left\{ -\frac{1}{M}\left(\lambda^{km}\sigma_{\rho}^{km}\right)^{2}-\nu^{k}P_{p}^{km}+{\displaystyle \sum_{n\in\mathbb{N}_{m}}}\mu^{kn}\left(I_{th}^{kn}-I^{kn}\right)\right.\nonumber \\
 & \left.\thinspace\thinspace\thinspace\thinspace\thinspace\thinspace\thinspace\thinspace\thinspace\thinspace+\beta^{km}\left(|h_{p}^{km}|^{2}P_{p}^{km}-\varrho\Omega|h_{p}^{k_{k}^{\prime}m}|^{2}P_{p}^{k_{k}^{\prime}m}\right)\right\} \label{eq:27}
\end{align}
From (\ref{eq:27}), it is clear that we can decompose the Lagrange
dual function $g^{k}\left(\mu,\nu,\beta\right)$ to $M$ independent
sub-problems by giving $\nu^{k}$. Each of the sub-problems is given
by
\begin{align}
\mathrm{OP2}: & \underset{P_{p}^{km}}{\max}-\frac{1}{M}\left(\lambda^{km}\sigma_{\rho}^{km}\right)^{2}-\nu^{k}P_{p}^{km}\\
\textrm{s.t.} & {\displaystyle {\displaystyle I^{kn}\leq I_{th}^{kn}}},\thinspace\thinspace n\in\mathbb{N}_{m}\label{eq:sub2.2}\\
 & |h_{p}^{km}|^{2}P_{p}^{km}\geq\varrho\Omega|h_{p}^{k_{k}^{\prime}m}|^{2}P_{p}^{k_{k}^{\prime}m}\label{eq:sub3.2}
\end{align}

The Lagrangian of OP2 is
\begin{align}
\mathcal{\tilde{L}}\left(\left\{ P_{p}^{km}\right\} ,\tilde{\mu}^{kn},\tilde{\beta}^{km}\right) & =-\frac{1}{M}\left(\lambda^{km}\sigma_{\rho}^{km}\right)^{2}-\nu^{k}P_{p}^{km}+{\displaystyle \sum_{n\in\mathbb{N}_{m}}}\tilde{\mu}^{kn}\left(I_{th}^{kn}-I^{kn}\right)\nonumber \\
 & \thinspace\thinspace\thinspace\thinspace\thinspace\thinspace\thinspace\thinspace\thinspace+\tilde{\beta}^{km}\left(|h_{p}^{km}|^{2}P_{p}^{km}-\varrho\Omega|h_{p}^{k_{k}^{\prime}m}|^{2}P_{p}^{k_{k}^{\prime}m}\right)\label{eq:L2}
\end{align}
where $\tilde{\mu}^{kn}$ and $\tilde{\beta}^{km}$ are the non-negative
dual variables for constraints (\ref{eq:sub2.2}) and (\ref{eq:sub3.2}),
respectively.

The Lagrange dual function of OP2 is given by
\begin{equation}
\tilde{g}^{km}\left(\tilde{\mu}^{kn},\tilde{\beta}^{km}\right)=\underset{P_{p}^{km}}{\max}\tilde{\mathcal{L}}\left(\left\{ P_{p}^{km}\right\} ,\tilde{\mu}^{kn},\tilde{\beta}^{km}\right)
\end{equation}
\textcolor{black}{The dual problem is then expressed as}
\begin{align}
\mathrm{min} & \tilde{g}^{km}\left(\tilde{\mu}^{kn},\tilde{\beta}^{km}\right)\\
\textrm{s.t.} & \tilde{\mu}^{kn}\geq0,\thinspace\forall n\in\mathbb{N}_{m}\\
 & \tilde{\beta}^{km}\geq0
\end{align}

\textcolor{black}{The opti}mal power allocation solution $\tilde{P}_{p}^{km}$
of OP2 can be obtained by using the Karush-Kuhn-Tucker (KKT) conditions
as (\ref{eq:P_end}) shows.
\begin{equation}
\tilde{P}_{p}^{km}=\underset{\textrm{geometric-dilution}}{\underbrace{\lambda^{km}}}\times\underset{\textrm{ranging-factor}}{\underbrace{\widetilde{\sigma}_{\rho}^{km}}}\times\underset{\textrm{constraint-scale}}{\underbrace{\left[M\left(\tilde{\beta}^{km}|h_{p}^{km}|^{2}-\nu^{k}-{\displaystyle \sum_{n\in\mathbb{N}_{m}}}\tilde{\mu}^{kn}J^{kn\leftarrow m}\right)\right]^{-1/2}}}\label{eq:P_end}
\end{equation}
where $J^{kn\leftarrow m}$ can be found in Appendix \ref{subsec:Derivation-of-1}.

\subsection{Remarks}

\textcolor{black}{From (\ref{eq:P_end}), it is observed that the
optimal power allocation solution is determined by the geometric-dilution,
ranging-factor and constraint-scale. It is necessary to have a clear
understanding of these factors that affect the allocated power.}

\textcolor{black}{The geometric-dilution $\lambda^{km}$ associates
with the relative positions between the P-User and all gNBs. This
means the power allocation procedure does not only minimize the ranging
accuracy, but also considers the geometric distribution which affects
the positioning accuracy as well.}

\textcolor{black}{The ranging-factor $\widetilde{\sigma}_{\rho}^{km}$
reflects the ranging ability of P-User as (\ref{eq:9}) shows. If
the loop parameters are fixed, $\widetilde{\sigma}_{\rho}^{km}$ is
determined by the channel gains of a certain P-User $m$, i.e. $\left|h_{p}^{km}\right|^{2}$,
$\left|h_{c}^{m\leftarrow k'}\right|^{2}$ and $\left|h_{p}^{k'm}\right|^{2}$
which reflect the attenuation of the positioning signal, the communication
signals and the other positioning signals, respectively. If the positioning
signal's attenuation is large, i.e. $\left|h_{p}^{km}\right|^{2}$
is small, it will allocate a stronger positioning power, and vice
versa. Conversely, if the communication signals' attenuation is large,
i.e. $\left|h_{c}^{m\leftarrow k'}\right|^{2}$ is small, it will
allocate a weak positioning power because of the small interference
from the communication signals, and vice versa. Please notice that
$\sum_{k'\in\mathcal{K}}\left|h_{c}^{m\leftarrow k'}\right|^{2}$
in $\widetilde{\sigma}_{\rho}^{km}$ will converge to $\left|h_{c}^{m\leftarrow k}\right|^{2}$
if the communication signals from all other gNBs (except gNB $k$)
are weak enough. However, different from $\left|h_{p}^{km}\right|^{2}$
and $\left|h_{c}^{m\leftarrow k'}\right|^{2}$, there is no monotonic
relation between the allocated power and $\left|h_{p}^{k'm}\right|^{2}$
as $\left|h_{p}^{k'm}\right|^{2}$ will be weighted by the powers
of the positioning signals broadcast by the other gNBs, i.e. $P_{p}^{k'm}$s.}

\textcolor{black}{The constraint-scale reflects the impact of the
constraints:}

\textcolor{black}{$\tilde{\mu}^{kn}$ is the dual variable associated
with the BER threshold of C-User $kn$. If  C-User $kn$ can accommodate
a higher BER, $\tilde{\mu}^{kn}$ will be smaller, and thus result
in a higher constraint-scale, and vice versa. In the extreme case
that C-User $kn$ cannot accommodate any additional interference,
$\tilde{\mu}^{kn}$ will be infinity, and thus the constraint-scale
will be zero, which indicates that the positioning signal over C-User
$kn$'s band is not permitted. On the contrary, if C-User $kn$ has
no requirement on the BER, $\tilde{\mu}^{kn}$ will be zero, and thus
the power-scale will be only determined by the other constraints.}

\textcolor{black}{$J^{kn\leftarrow m}$ is determined by the channel
gains of positioning signals $k'm$ ($k'\in\mathcal{K}$) at C-User
$kn$ as (\ref{eq:56}) shows. It is clear that a smaller $\sum_{k'\in\mathcal{K}}\left|h_{p}^{kn\leftarrow k'm}\right|^{2}$will
result in a higher constraint-scale. This is intuitively correct because
the P-Users' transmissions from all gNBs will not cause too much interference
when $\sum_{k'\in\mathcal{K}}\left|h_{p}^{kn\leftarrow k'm}\right|^{2}$
is small. In the real scenarios, if $\sum_{k'\in\mathcal{K}}\left|h_{p}^{kn\leftarrow k'm}\right|^{2}\rightarrow0$,
which means C-User $kn$ is too far from all gNBs to receive any communication/positioning
signal, the P-Users will not cause any interferences to this C-User
no matter how strong its transmit power is. If C-User $kn$ is far
from all gNBs except its own cell, i.e. $\sum_{k'\in\mathcal{K}}\left|h_{p}^{kn\leftarrow k'm}\right|^{2}\rightarrow h_{p}^{kn\leftarrow km}\approx h_{c}^{kn}$,
the positioning signals from other gNBs will not interference the
C-Users.}

\textcolor{black}{$\nu^{k}$ is the dual variable associated with
the transmit power budget. A larger power budget results in a smaller
$\nu^{k}$, and thus results in a lower ranging error, and vice versa.}

\textcolor{black}{$\tilde{\beta}^{km}$ is a parameter related to
the P-User's receiver performance. It reflects the influence of the
cross-correlation (i.e. (\ref{eq:sub_3})) on the constraint-scale.
There will be a higher $\tilde{\beta}^{km}$ with a smaller $\varrho$.
Namely, if the receiver has a better anti-cross-correlation performance,
there will be a higher ranging accuracy and better coverage, vice
versa.}

\subsection{The Positioning-Communication Joint Power Allocation Scheme}

\textcolor{black}{The remaining task for solving OP1 is to obtain
the optimal dual variables, which are the same in both OP1 and OP2.
Applying the solution to OP2, we can obtain the optimal power allocation
$\tilde{P}_{p}^{km}$ in OP1. However, it is difficult to solve OP2
directly because we cannot obtain the closed-form expression for dual
variables. }The Lagrange dual function\textcolor{black}{{} (\ref{eq:Fen_1})
}is made up of $K$ independent sub-problems.\textcolor{black}{{} For
each }sub-problem\textcolor{black}{, it is observed that $\nu^{k}$
is the same for all P-Users. $\mu^{kn}$ and $\beta^{km}$ are different
for C-Users and P-Users, respectively. Then, we can solve the optimization
problem using hierarchical algorithm by updating the values of the
dual variables $\left\{ \mu,\nu,\beta\right\} $ via subgradient methods,
which guarantees the gradient-type algorithm to converge to the optimal
solution \cite{6775304}.}

\textbf{\textcolor{black}{Proposition}}\textcolor{black}{: The subgradients
of $\tilde{g}^{km}\left(\tilde{\mu}^{kn},\tilde{\beta}^{km}\right)$
are given by $\grave{\mu}^{kn}=I_{th}^{kn}-I^{kn}$ and $\grave{\beta}^{km}=|h_{p}^{km}|^{2}P_{p}^{km}-\varrho\Omega|h_{p}^{k_{k}^{\prime}m}|^{2}P_{p}^{k_{k}^{\prime}m}$.
Then, $\tilde{P}_{p}^{km}$ is the optimal solution obtained at $\tilde{\mu}^{kn}$
and $\tilde{\beta}^{km}$. Next, $\nu^{k}$ is updated by its subgradient,
which is given by $\grave{\nu}^{k}=P_{th}^{k}-{\displaystyle \sum_{m\in\mathcal{M}}P_{p}^{km}}$.
Finally, $\hat{P}_{p}=\left\{ \tilde{P}_{p}^{km},\forall k\in\mathcal{K},\forall m\in\mathcal{M}\right\} $
is the optimal solution obtained at $\nu^{k}$ under the given $\mu^{kn}$
and $\beta^{km}$.}

\textbf{\textcolor{black}{Proof}}\textcolor{black}{: Please see the
Appendix \ref{subsec:Subgradient-method-of}.}

\textcolor{black}{Using the above gradient, we can obtain the optimal
power allocation by iterative Lagrangian multiplier $\left\{ \mu,\nu,\beta\right\} $.
Notice that the positions and powers of P-Users are unknown which
are necessary for calculating the geometric-dilution $\lambda^{km}$
and the ranging factor $\widetilde{\sigma}_{\rho}^{km}$, respectively.
We can minimize the ranging error for all positioning users without
considering the impact of the geometric-dilution and the impact of
multiple access at the first iteration. Then, we can get approximate
positions and initial allocated powers. After several iterations,
the positions and the powers will be converge to the optimal values.}

\textcolor{black}{The algorithm to solve OP1 can be summarized as
Algorithm \ref{alg:The-sub-gradient-Method} shows. Where $t$ and
$t'$ are the iteration numbers. $iterN$ is the maximum iteration
amount. $b_{1}$, $b_{2}$ and $b_{3}$ are the update step sizes.
$\varepsilon>0$ is a given small constant.}

\begin{algorithm}[tbh]
\caption{\textcolor{black}{The proposed PCJPA }algorithm\label{alg:The-sub-gradient-Method}}

\begin{algorithmic}[1]

\STATE Initial the dual variable $\nu_{1}^{k}$ for all $k\in\mathcal{K}$
\textbf{in parallel}

\FOR{$t=1\textrm{ to }iterN$}

\STATE Initial $\mu_{1}^{kn},\beta_{1}^{km},P_{p}^{k'm}$ for all
$n\in\mathcal{N}$, \textbf{$m\in\mathcal{M}$ }and\textbf{ $k'\in\mathcal{K}^{k}$
in parallel}

\FOR{$t'=1\textrm{ to }iterN$}

\STATE For each P-User $m$, calculate $\tilde{P}_{p}^{km}$ using
(\ref{eq:P_end})

\STATE Update $\mu_{t'}^{kn}$ and $\beta_{t'}^{km}$ by their subgradient:

\STATE i) $\tilde{\mu}_{t'+1}^{kn}=\mu_{t'}^{kn}-b_{2}\grave{\mu}^{kn}$

\STATE ii) $\tilde{\beta}_{t'+1}^{km}=\beta_{t'}^{km}-b_{3}\grave{\beta}^{km}$

\STATE Update $P_{p}^{km}$=$\tilde{P}_{p}^{km}$ for all $k\in\mathcal{K}$
and $m\in\mathcal{M}$

\IF{ $|\tilde{\mu}_{t'+1}^{kn}-\mu_{t'}^{kn}|\leq\varepsilon$ \&
$|\tilde{\beta}_{t'+1}^{km}-\beta_{t'}^{km}|\leq\varepsilon$}

\STATE break

\ENDIF   

\ENDFOR

\STATE Update $\nu_{t+1}^{k}$ by its subgradient:

$\tilde{\nu}_{t+1}^{k}=\nu_{t}^{k}-b_{1}\grave{\nu}^{k}$

\IF{ $|\tilde{\nu}_{t+1}^{k}-\nu_{t}^{k}|\leq\varepsilon$}

\STATE break

\ENDIF   

\ENDFOR

\RETURN $\hat{P}_{p}=\left\{ \tilde{P}_{p}^{km},\forall k\in\mathcal{K},\forall m\in\mathcal{M}\right\} $

\end{algorithmic} 
\end{algorithm}

\section{The Performance Evaluation\label{sec:Performance-evaluation}}

In this section, we evaluate the communication and ranging performances
of the proposed MS-NOMA signal under a single cell network firstly.
Then, we examine the positioning performance in a 4-gNBs cell by considering
the impact of all factors comprehensively. In the single cell scenario,
the impact of the cross-correlation will be vanished and the channel
gains of the communication and positioning signals will be seen as
equal. Then, (\ref{eq:1}) and (\ref{eq:7-1}) become
\begin{equation}
BER^{n}=\Gamma\textrm{erfc}\left(\frac{\gamma P_{c}T_{c}}{I^{n}+2N_{0}}\right)\label{eq:36}
\end{equation}
\textcolor{black}{
\begin{align}
\left(\sigma_{\rho}^{m}\right)^{2} & \approx\frac{aT_{p}^{2}}{2}\left[\frac{1}{B_{fe}T_{p}\left(C/N_{0}\right)^{m}}+\frac{B\left(CPR\right)^{m}}{2B_{fe}^{2}}\right]\label{eq:37}
\end{align}
where
\begin{align}
I^{n} & =\sum_{m=1}^{M}P_{p}^{m}T_{p}\textrm{sinc}^{2}\left(m-\frac{n}{G}\right)
\end{align}
\begin{equation}
\left(C/N_{0}\right)^{m}=P_{p}^{m}/N_{0}
\end{equation}
\begin{equation}
\left(CPR\right)^{m}=2GP_{c}/P_{p}^{m}
\end{equation}
Notice that (\ref{eq:36}) and (\ref{eq:37}) reflect the features
of MS-NOMA signal itself without interferences from other gNBs.}

The simulation settings are: The communication and positioning signals
use QPSK and BPSK constellation, respectively. \textcolor{black}{The
carrier frequency is set to 3.5GHz and $\varDelta f_{c}=30\textrm{kHz}$.}
Two scenarios with different bandwidths of positioning signals are
considered: 1) $B=20\textrm{MHz}$ 2) $B=50\textrm{MHz}$. The amount
of\textcolor{black}{{} P-User is $M=20$, i.e. }the positioning signal
is 30/80 times faster than the communication one under 20/50MHz bandwidth,
respectively.\textcolor{black}{{} The powers of all C-Users are assumed
as identical. The front-end bandwidth is set to twice of the total
bandwidth, i.e. $B_{fe}=2B$. The loop parameters are set as: $B_{L}=0.2\textrm{Hz}$,
$T_{coh}=0.02s$ and $D=0.02\textrm{chips}$, where $B_{L}$ is the
code loop noise bandwidth and $T_{coh}$ is the predetection integration
time.}

\subsection{Communication\textcolor{black}{{} Performance}}

\begin{figure}[tbh]
\begin{centering}
\subfloat[\textcolor{black}{Average BER in different scenarios\label{fig:Average-BER}}]{\begin{centering}
\includegraphics[viewport=0bp 0bp 370bp 315bp,clip,width=0.45\columnwidth]{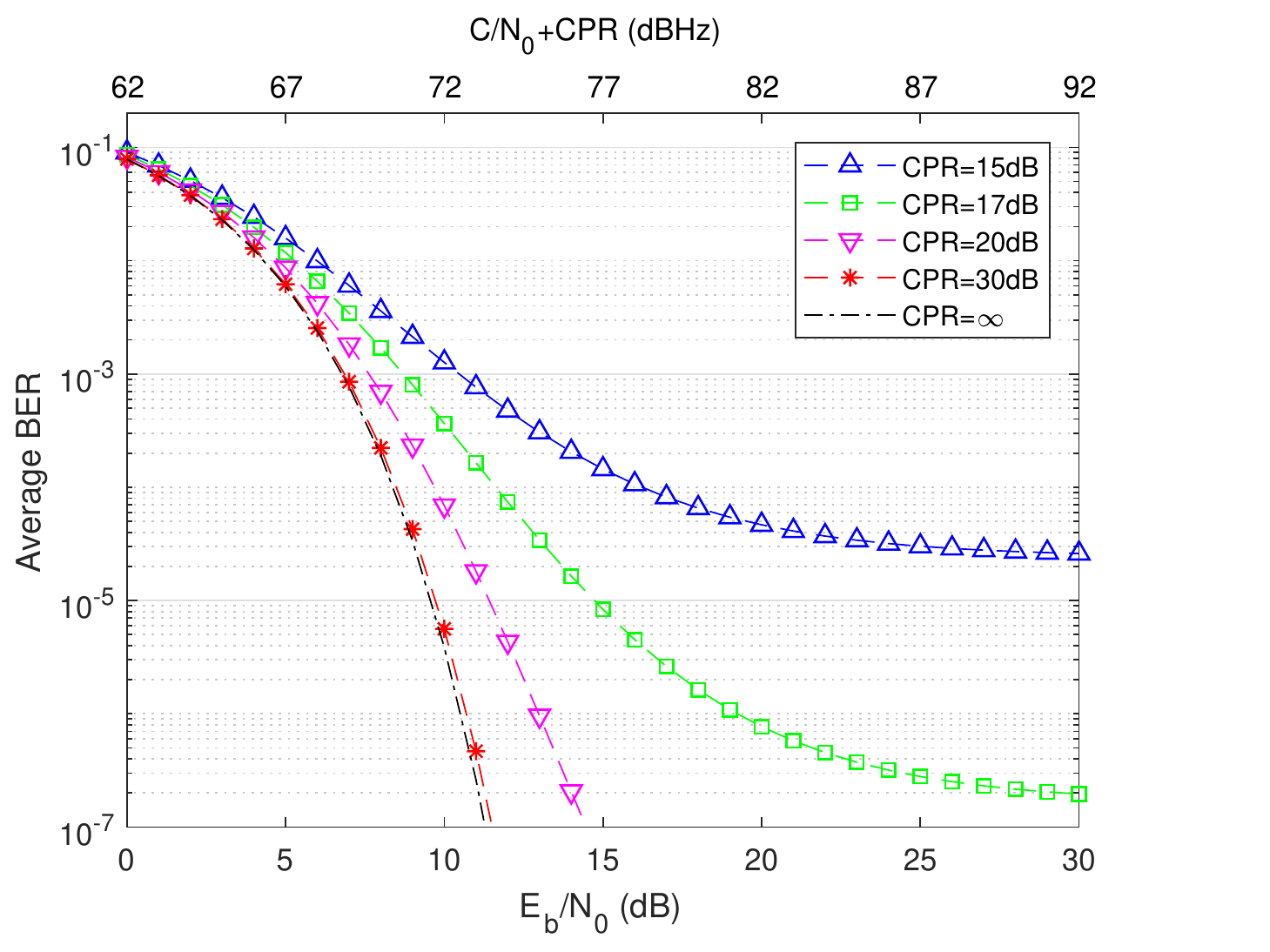}
\par\end{centering}
}\subfloat[\textcolor{black}{BER of the communication signal. $E_{b}/N_{0}=5\textrm{dB}$,
$CPR=15\textrm{dB}$\label{fig:BER-of-the}}]{\begin{centering}
\includegraphics[width=0.45\columnwidth]{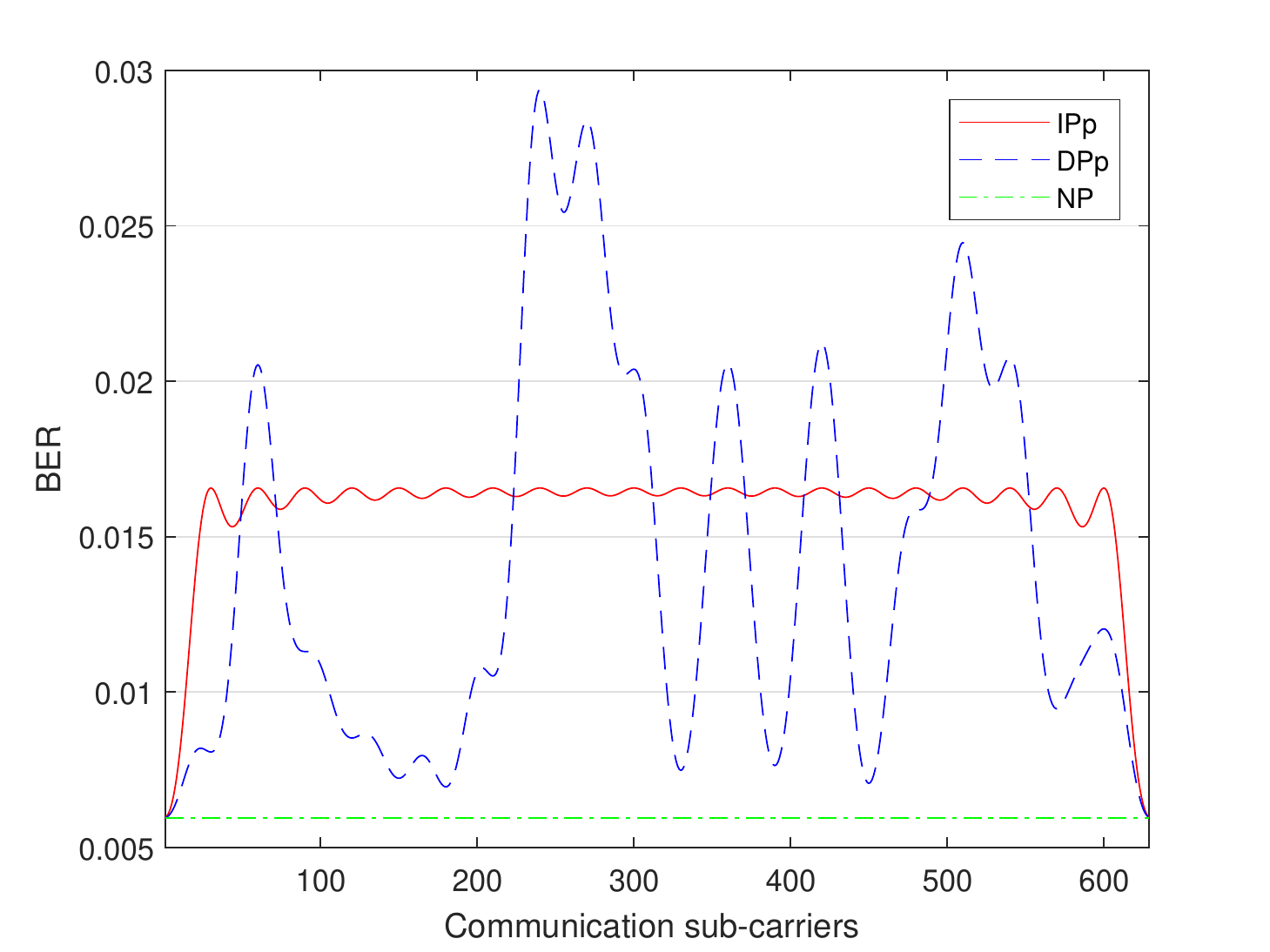}
\par\end{centering}
}
\par\end{centering}
\caption{Communication\textcolor{black}{{} Performance\label{fig:Communication-Performance}}}
\end{figure}

We firstly examine the i\textcolor{black}{nterference of the positioning
signal to the C-Users. Fig. \ref{fig:Average-BER} shows the average
BERs over the whole bandwidth when $P_{p}^{m}=P_{p},\thinspace\forall m\in\mathcal{M}$.
It is clear that the average BERs decrease with the increasing of
$E_{b}/N_{0}$ as well as $C/N_{0}$ ($E_{b}=P_{c}T_{c}$ is the energy
of the communication symbol). Notice that the BER curves with small
CPR will tend to be flat when $E_{b}/N_{0}$ becomes larger. This
is because the interference caused by the positioning signal dominates
the BER performance rather than the environment noise (i.e. $I^{n}$
is much larger than $2N_{0}$). When the positioning signal becomes
weaker (CPR becomes larger), the BER curves will become flat with
larger $E_{b}/N_{0}$ and they will become closer to the one that
only exists noise ($\textrm{CPR}=\infty$).}

\textcolor{black}{Fig. \ref{fig:BER-of-the} detailed shows the BERs
for each C-User. If the powers of the P-Users are identical (IPp),
the BERs are approximately identical as well. While the BERs are different
when the P-Users' powers are different (DPp). Then, the maximum BER
is related to all P-Users' powers (see (\ref{eq:36})) in this case.
Of course, both of the IPp and DPp are higher than the scenario that
do not exist positioning signals (NP).}

\subsection{Ranging \textcolor{black}{Performance}}

We then exa\textcolor{black}{mine the range measurement accuracy of
the MS-NOMA signal. The ranging accuracy of the MS-NOMA and PRS signal
are compared. Where the lower bound of PRS is used as introduced in
\cite{Peral2012Joint}.}
\begin{center}
\begin{figure}[tbh]
\begin{centering}
\includegraphics[viewport=0bp 0bp 370bp 315bp,clip,width=0.5\columnwidth]{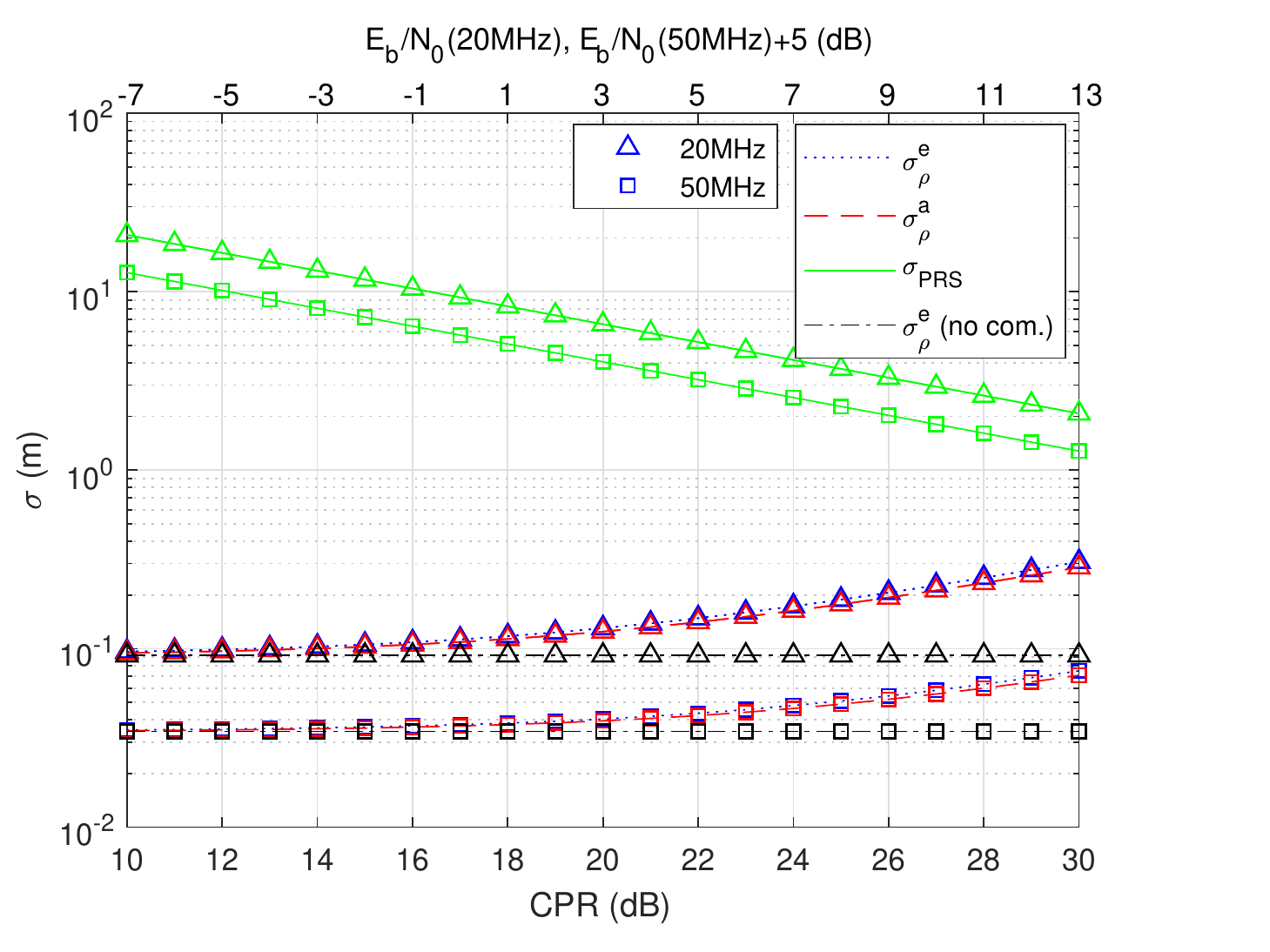}
\par\end{centering}
\caption{\textcolor{black}{Range measurement accuracy at $C/N_{0}=45\textrm{dB}\cdot\textrm{Hz}$\label{fig:Range-measurement-accuracy-1-1}}}
\end{figure}
\par\end{center}

\textcolor{black}{Fig. \ref{fig:Range-measurement-accuracy-1-1} shows
the range measurement accuracy when $C/N_{0}=45\textrm{dB}\cdot\textrm{Hz}$.
Where superscript }\textit{\textcolor{black}{e}}\textcolor{black}{{}
and }\textit{\textcolor{black}{a}}\textcolor{black}{{} represent the
exact and approximate results, respectively. It is clear that the
measurement errors of the MS-NOMA signal are always smaller than the
ones of the PRS signal when CPR<30dB. And the ranging accuracy of
the 20MHz MS-NOMA signal is even much higher than the one of the 50MHz
PRS signal. The accuracy gap between the two signals becomes larger
when CPR decreases. It is because when the power of the communication
signal decreases, its interference to the positioning one becomes
weaker. While the measurement accuracy of PRS becomes worse due to
the lower signal-to-noise ratio of the communication signal.}

\textcolor{black}{Please notice that the curves of the MS-NOMA signal
will not change if the communication power is fixed and the positioning
one is variable. For example, if $E_{b}/N_{0}=0$dB and $C/N_{0}$
varies from 32 to 52dB$\cdot\textrm{Hz}$ (i.e. 10dB<CPR<30dB) with
20MHz bandwidth, the measurement accuracy of PRS is fixed at 9.29m
and the accuracy of MS-NOMA is the same as Fig. \ref{fig:Range-measurement-accuracy-1-1}
shows. Then, the measurement error will decreases when the power of
the positioning signal increases.}

\textcolor{black}{Although stronger positioning power will have higher
measurement accuracy, the maximum power of the positioning signals
will be limited by the QoS of communication as Fig. \ref{fig:Communication-Performance}
shows. So, $P_{p}^{m}$s must be allocated carefully to acquire the
best ranging performance under the QoS constraint. In the real application,
both of the communication and positioning interferences from other
gNBs, must be considered which will be evaluated next subsection.}

\textcolor{black}{Fig. \ref{fig:Range-measurement-accuracy-1-1} also
confirms that the approximations of $\sigma_{\rho}$ (see (\ref{eq:37}))
correspond to the exact one (see (\ref{eq:3})) very well. And all
of them are decimeter or centimeter level which ensures sub-meter
level positioning accuracy compare to the meter level positioning
accuracy of PRS. Meanwhile, the measurement error is o}nly a little
larger than the one that without communication signal, which means
the effect of the communication signal is limit to the range measuring.

\subsection{Positioning \textcolor{black}{Performance}\label{sec:Numerical-results}}

In this section, we present the numerical results to evaluate the
positioning \textcolor{black}{performance} of MS-NOMA signal by using
the proposed PCJPA algorithm. The gNBs are fixed at $\left(0,0\right)$,
$\left(0,200\right)$, $\left(200,200\right)$, $\left(200,0\right)$
and 20\textcolor{black}{{} P-Users are randomly distributed in the coverage
area.} The free space propagation model is employed with 50 Monte
Carlo runs in each simulation. \textcolor{black}{Without any loss
of generality, we set $P_{th}^{k}=P_{th}$ for any $k\in\mathcal{K}$
and $\varrho=2$.}

\begin{figure}[tbh]
\centering{}\subfloat[\textcolor{black}{The proposed PCJPA algorithm\label{fig:the-proposed-joint}}]{\begin{centering}
\includegraphics[width=0.45\columnwidth]{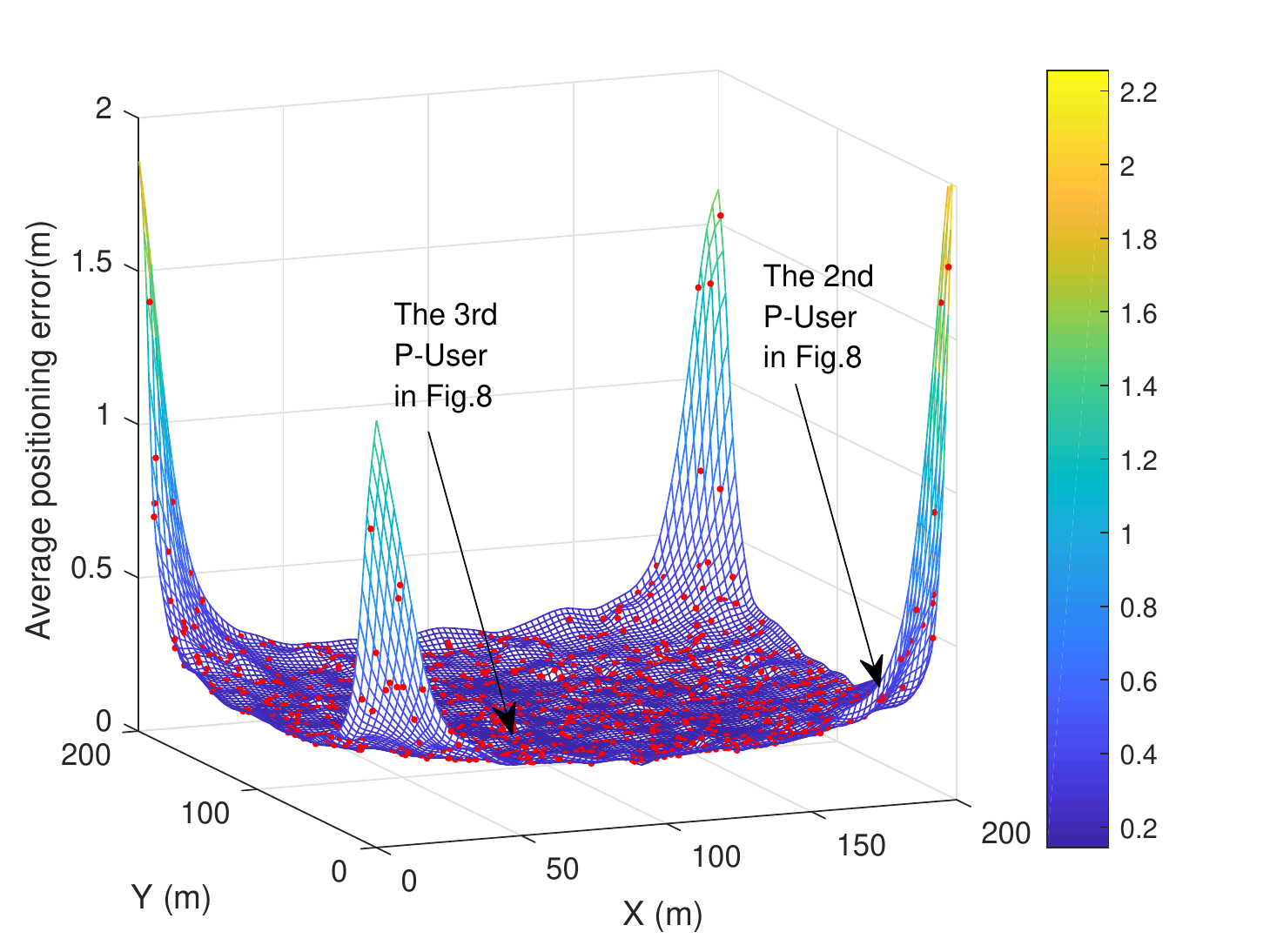}
\par\end{centering}
}\subfloat[The traditional method without power allocation\label{fig:The-traditional-method}]{\begin{centering}
\includegraphics[width=0.45\columnwidth]{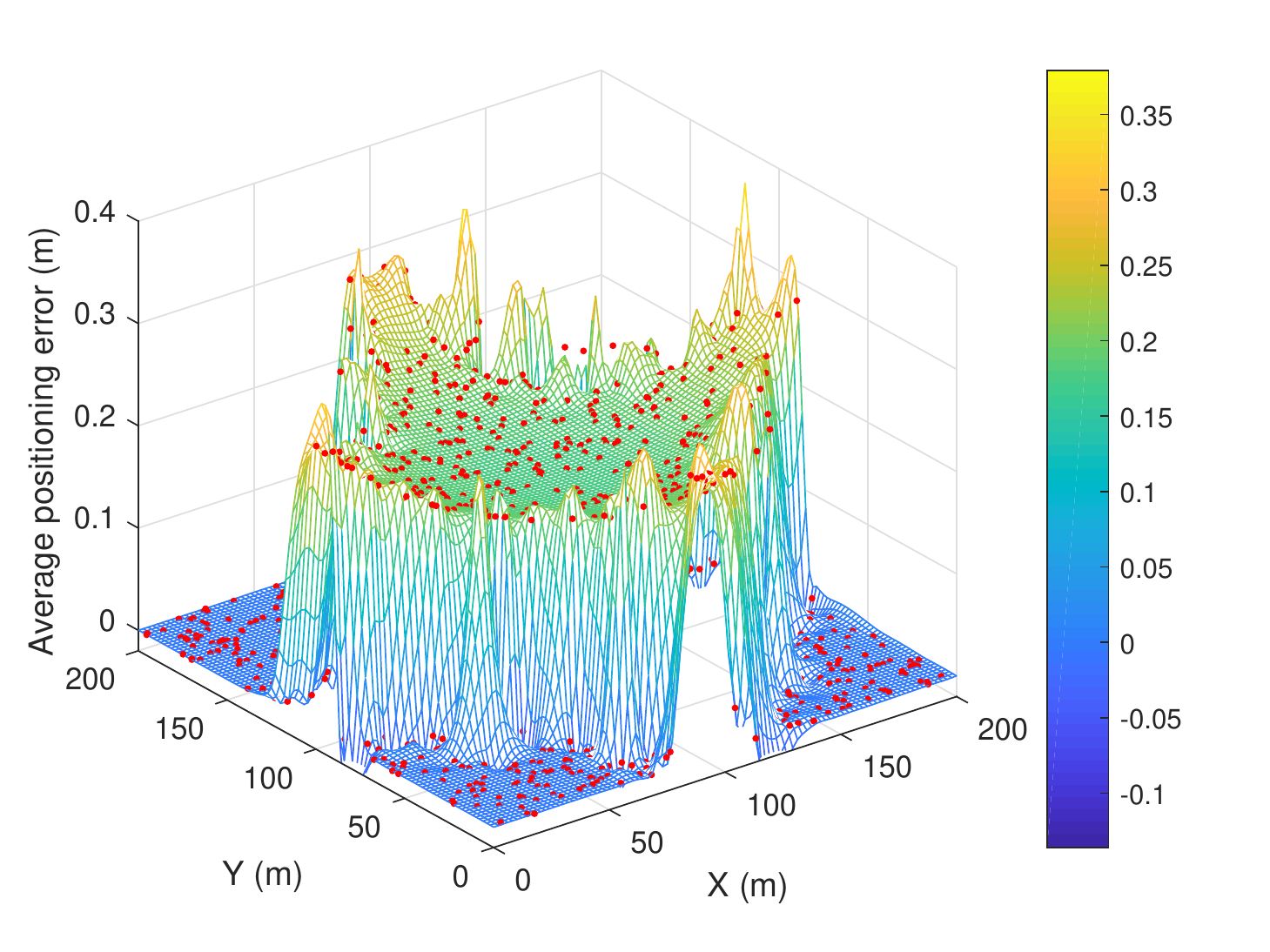}
\par\end{centering}
}\caption{The\textcolor{black}{{} }positioning signal's coverage and positioning
accuracy ($B=50\textrm{MHz}$, $P_{th}=0.8$W, $\beta_{th}=8\times10^{-3}$)\label{fig:Relitu}}
\end{figure}

\textcolor{black}{Fig. \ref{fig:Relitu} shows the }positioning signal's
coverage and positioning accuracy\textcolor{black}{{} }of MS-NOMA signal
\textcolor{black}{by using the proposed PCJPA algorithm and the traditional
equal-power transmission strategy, respectively. The positioning error
is set to 0 if the P-User can not receive more than 3 gNBs which means
there are no enough gNBs to position. From Fig. \ref{fig:the-proposed-joint},
we can see all P-Users have positioning results which means the near-far
effect is dramatically reduced. While from Fig. \ref{fig:The-traditional-method},
it is clear that a great amount of P-Users (50.8\%) do not have positioning
results by the traditional method as suffering severe near-far effect.}

Detailed comparisons are presented in Table \ref{tab:The-comparison-of}.
The positioning accuracy of MS-NOMA signal is much higher than the
one of PRS. Specifically, the improvements of MS-NOMA signal (using
the proposed PCJPA algorithm) are 93.7\% and 93.1\% compared to PRS
with 20MHz and 50MHz bandwidth, respectively. Notice that the positioning
error of PCJPA is a little higher than the equal-power strategy. This
is because the\textcolor{black}{{} edge of the coverage area have poor
geometric distribution which lead to poor positioning accuracy, while
these low accuracy P-Users are not averaged in the equal-power strategy
as they have no positioning results. In fact, the positioning errors
of the PCJPA algorithm are smaller than the ones of the equal-power
strategy by examining the same area as Fig. \ref{fig:The-CDF-of}
shows (0.17m v.s. 0.20m at 50MHz (15\% improvement) and 0.41m v.s.
0.54m at 20MHz (24\% improvement)). Only 10\% P-Users' positioning
errors by PCJPA algorithm are larger than the ones by equal-power
strategy. While more than 50\% P-Users do not have positioning results
if power allocation is not executed. Therefore, the proposed PCJPA
algorithm has excellent performances in both coverage and positioning
accuracy than the traditional scheme.}

\begin{table}[tbh]
\caption{The comparison between different signals by using different power
allocation strategies\label{tab:The-comparison-of}}

\centering{}%
\begin{tabular}{|c|c|c|c|c|}
\hline 
\multicolumn{2}{|c|}{Signal} & \multicolumn{2}{c|}{MS-NOMA} & PRS\tabularnewline
\hline 
\multicolumn{2}{|c|}{Power allocation strategy} & PCJPA & Equal power & Equal power\tabularnewline
\hline 
\hline 
\multirow{2}{*}{Positioning Error} & 20MHz & 0.57m & 0.54m & 8.44m\tabularnewline
\cline{2-5} 
 & 50MHz & 0.21m & 0.20m & 3.17m\tabularnewline
\hline 
\end{tabular}
\end{table}

\begin{center}
\textcolor{black}{}
\begin{figure}[tbh]
\begin{centering}
\includegraphics[width=0.5\columnwidth]{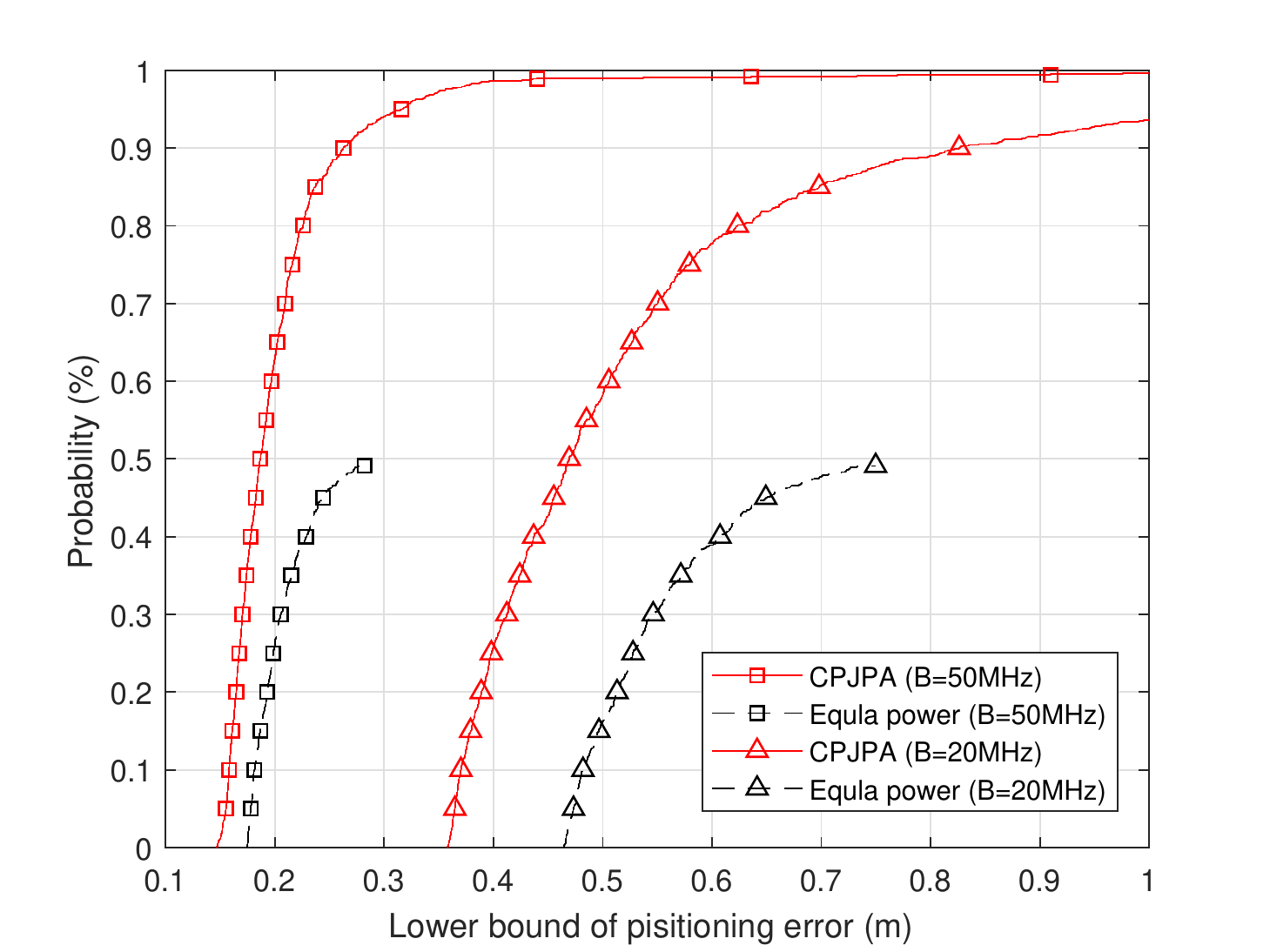}
\par\end{centering}
\textcolor{black}{\caption{\textcolor{black}{The CDF of the MS-NOMA signal with different power
allocation strategies }($P_{th}=0.8$W, $\beta_{th}=8\times10^{-3}$)\textcolor{black}{\label{fig:The-CDF-of}}}
}
\end{figure}
\par\end{center}

\begin{figure}[tbh]
\begin{centering}
\subfloat[Average positioning errors under different QoS requirements\label{fig:Average-measurement-error}]{\begin{centering}
\includegraphics[width=0.45\columnwidth]{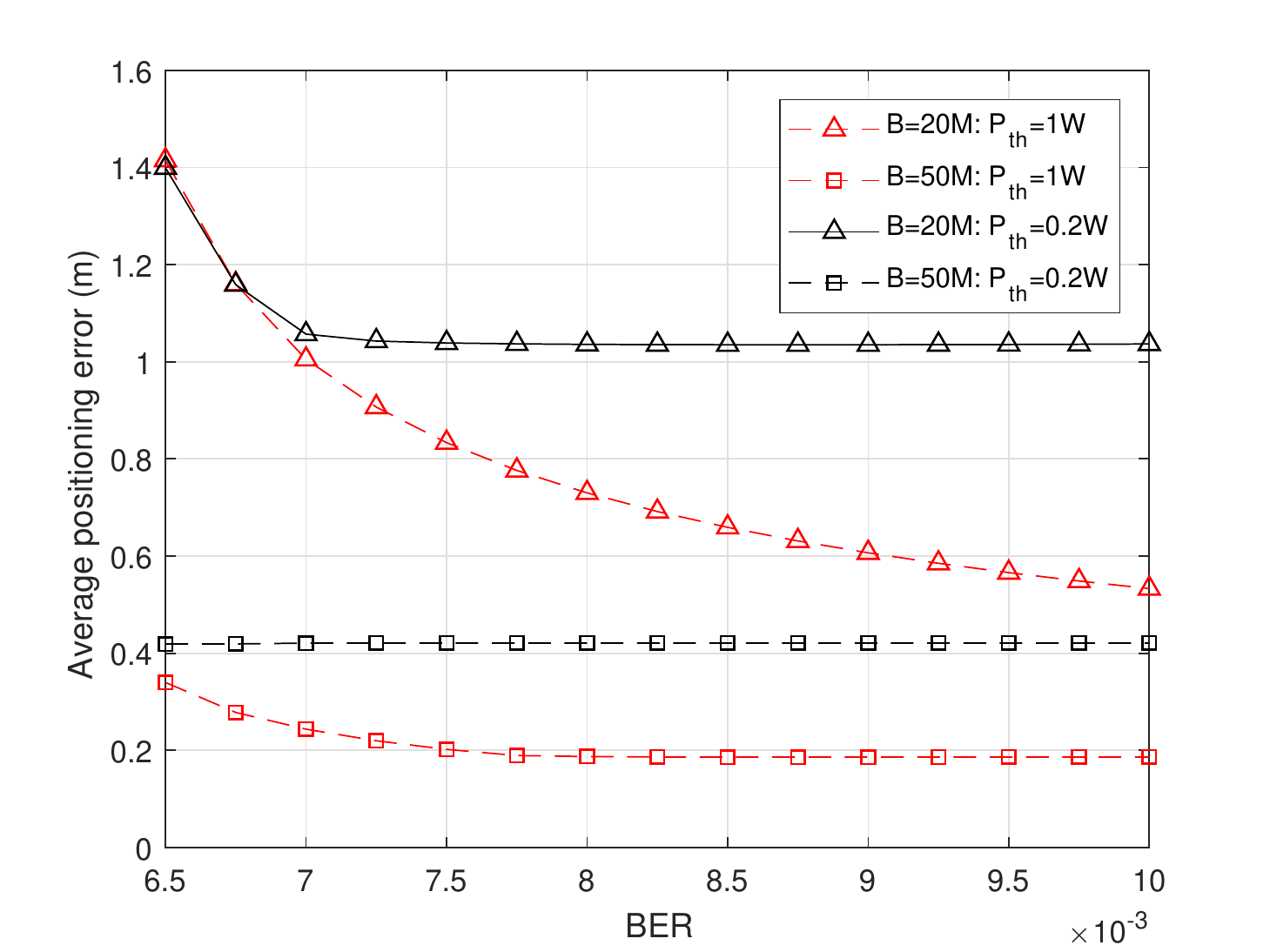}
\par\end{centering}
\centering{}}\subfloat[Average positioning errors under different power budgets\label{fig:Average-measurement-error-1}]{\begin{centering}
\includegraphics[width=0.45\columnwidth]{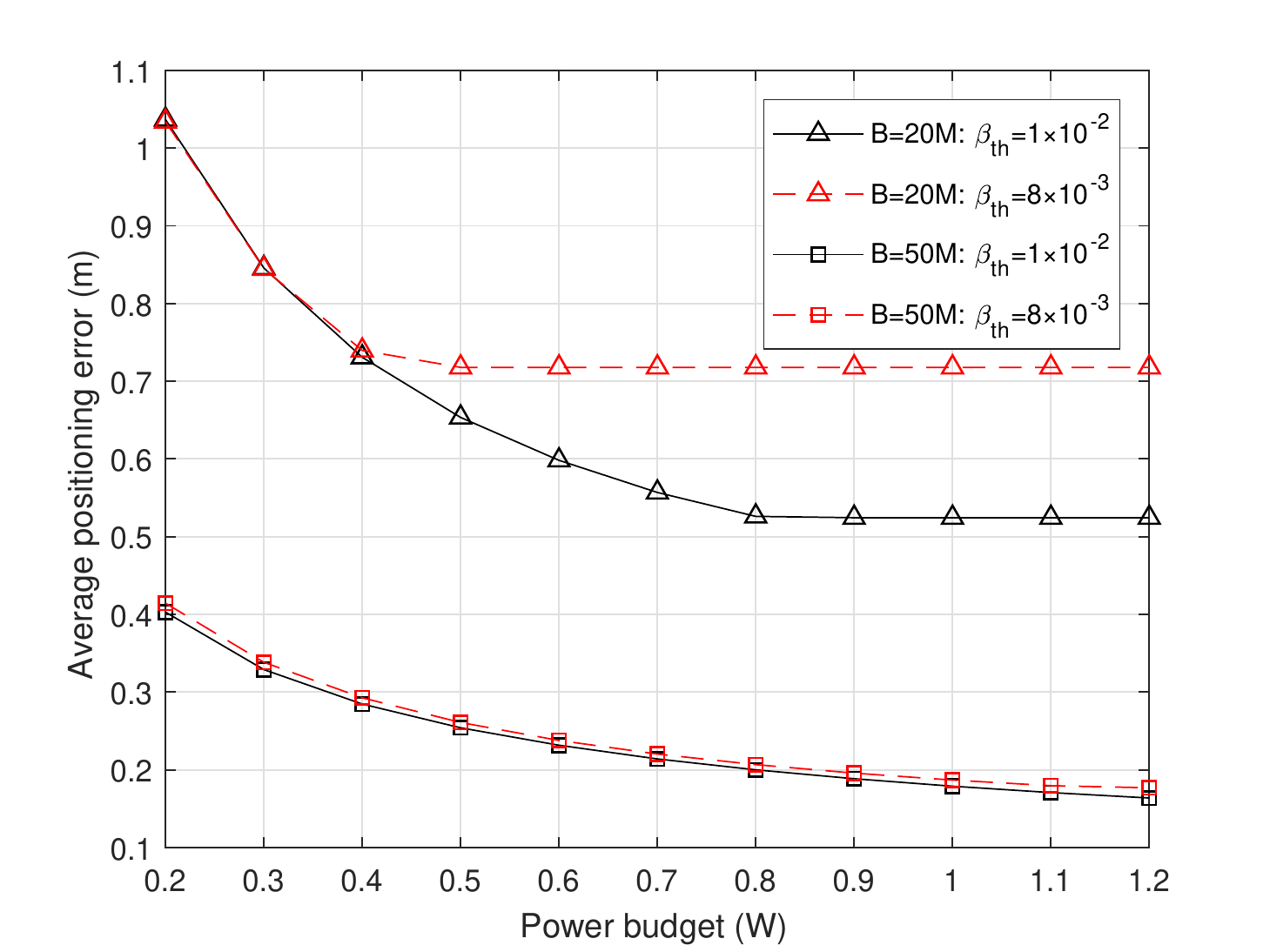}
\par\end{centering}
\centering{}}
\par\end{centering}
\caption{The impact of different constraints and parameters\label{fig:The-impact-of}}
\end{figure}

Fig. \ref{fig:The-impact-of} illustrated the impact of different
constraints and parameters on the positioning accuracy. From Fig.
\ref{fig:Average-measurement-error}, it is clear that the average
positioning errors decrease with the increasing of the tolerable BER.
Meanwhile, there is higher positioning accuracy with larger bandwidth.
Notice that the curves tend to constant values when the tolerable
BER increases. It is because the power budget limits the improvement
of the performance. \textcolor{black}{And it is clear that higher
power budget can obtain a lower positioning error bound. }On the other
hand, the curve with a smaller bandwidth converges slowly.\textcolor{black}{{}
It is because the power of the positioning signal with a small bandwidth
will concentrated in a narrower range. Consequently, there will be
more powers of P-Users leak to the neighbor C-Users as (\ref{eq:2})
and (\ref{eq:4}) show which leads to more interferences.}

Fig.\textcolor{black}{{} \ref{fig:Average-measurement-error-1} shows
that the average }positioning\textcolor{black}{{} error decreases with
the increasing of the power budget. Similar to }Fig.\textcolor{black}{{}
\ref{fig:Average-measurement-error}, the curves tend to }constant\textcolor{black}{{}
values when the power budget increases as well. This means the QoS
constraint becomes dominant and the average }positioning\textcolor{black}{{}
errors do not decrease although the total power increases. If we have
a strict QoS constraint (smaller $\Xi_{th}$), the positioning error
bound will be higher.}

\begin{figure}[tbh]
\begin{centering}
\includegraphics[width=0.5\columnwidth]{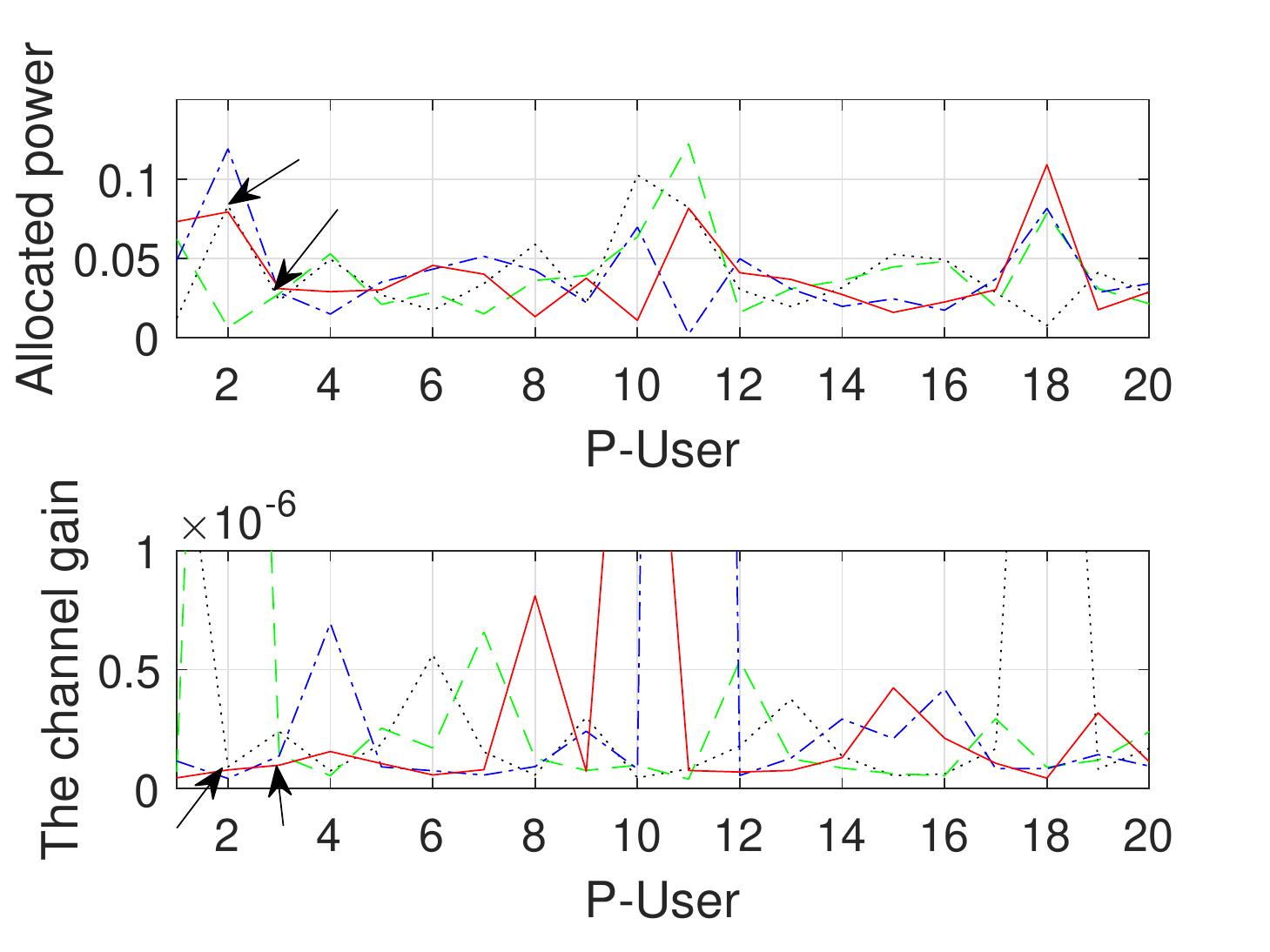}
\par\end{centering}
\centering{}\caption{The allocated powers with different channel gains (\textcolor{black}{Different
lines represent different gNBs}) ($B=50\textrm{MHz}$, $P_{th}=0.8$W,
$\beta_{th}=8\times10^{-3}$)\label{fig:power-allocation}}
\end{figure}

\textcolor{black}{Fig. \ref{fig:power-allocation} illustrates the
relationship between the allocated powers and the channel gains of
the P-Users by examining one simulation.} It shows that the worse
channel states tend to allocate stronger positioning signals.\textcolor{black}{{}
This is exactly what we expected that the P-Users with worse channel
states need stronger powers to obtain an accurate ranging. }However,
notice that there is a power disparity between the $2^{\textrm{ed}}$
and $3^{\textrm{rd}}$ P-Users whose channel gains are similar. This
is because the geometric-dilution $\lambda$ in (\ref{eq:1-2}) also
affects the positioning accuracy which is considered by the power
allocation process. This can be observed in Fig. \ref{fig:the-proposed-joint}
as well: The location of the $2^{\textrm{ed}}$ P-User (coordinate:
$\left(178,8\right)$) is at the edge of the area compared to the
center location of the $3^{\textrm{rd}}$ P-User (coordinate: $\left(79,77\right)$),
so the former one has a stronger allocated power.

\section{Conclusions\label{sec:Conclusions}}

In this paper, we presented a feasibility study for a novel positioning-communication
integrated signal called Multi-Scale Non-Orthogonal Multiple Access
(MS-NOMA) for 5G positioning. The MS-NOMA signal superposes power
configurable positioning signals on the communication ones to achieve
high ranging accuracy and excellent signal coverage (less \textcolor{black}{near-far
effect}). \textcolor{black}{Like the normal NOMA signal, there are
interferences between the commutation and positioning signals. So,
we analyzed the BER for communication and the ranging error for positioning
when two kinds of signals exist simultaneously. The results show the
interaction is rather limited and the proposed MS-NOMA signal greatly
improves the ranging accuracy than traditional 5G signal. In addition,
because the positioning signals in the proposed MS-NOMA are power
controllable and multiple accessible, we modeled a multi-user power
allocation problem for an optimal positioning accuracy and signal
coverage as a convex optimization problem under the QoS requirement
and other constraints. Then, a novel Positioning-Communication Joint
Power Allocation (PCJPA) algorithm was proposed for solving this problem.
The numerical results show our proposed MS-NOMA signal with PCJPA
algorithm improves the positioning accuracy and signal coverage dramatically
than the traditional 5G signal with equal power transmission strategy.}

\appendix{}

\subsection{Derivation of the Horizontal Positioning Accuracy\label{subsec:Derivation-of-the}}

Define $\varepsilon_{\rho}^{m}=\left[\varepsilon_{\rho}^{1m},\varepsilon_{\rho}^{2m},...,\varepsilon_{\rho}^{km}\right]^{T}$
as the ranging errors of P-User $m$, where $\varepsilon_{\rho}^{km}$
represents the ranging error between gNB $k$ and P-User $m$. Then,
the positioning error of P-User $m$ is \cite{Lu2016Quality}
\begin{align}
\varepsilon_{X}^{m} & =\left[\left(G^{m}\right)^{T}G^{m}\right]^{-1}\left(G^{m}\right)^{T}\varepsilon_{\rho}^{m}\nonumber \\
 & =H^{m}\varepsilon_{\rho}^{m}\label{eq:1-1}
\end{align}
where
\begin{equation}
G^{m}=\left[\begin{array}{ccc}
\iota_{x}^{1m} & \iota_{y}^{1m} & \iota_{z}^{1m}\\
\iota_{x}^{2m} & \iota_{y}^{2m} & \iota_{z}^{2m}\\
\cdots & \cdots & \cdots\\
\iota_{x}^{km} & \iota_{y}^{km} & \iota_{z}^{km}
\end{array}\right]
\end{equation}
\begin{equation}
\left\{ \begin{array}{c}
\iota_{x}^{km}=\nicefrac{\left(x_{p}^{m}-x_{b}^{k}\right)}{\left\Vert X_{b}^{k}-X_{p}^{m}\right\Vert }\\
\iota_{y}^{km}=\nicefrac{\left(y_{p}^{m}-y_{b}^{k}\right)}{\left\Vert X_{b}^{k}-X_{p}^{m}\right\Vert }\\
\iota_{z}^{km}=\nicefrac{\left(z_{p}^{m}-z_{b}^{k}\right)}{\left\Vert X_{b}^{k}-X_{p}^{m}\right\Vert }
\end{array}\right.
\end{equation}
where $X=\left[x,y,z\right]^{T}$ represents the coordinate. Subscript
$p$ and $b$ represent P-User and gNB, respectively. Because the
ranging errors from the gNBs are independent, their covariance matrix
is diagonal under the assumption that the range measuring is unbiased:
\begin{align}
\left(\sigma_{\rho}^{m}\right)^{2} & =\textrm{cov}\left(\varepsilon_{\rho}^{m},\varepsilon_{\rho}^{m}\right)\nonumber \\
 & =\left[\begin{array}{cccc}
\left(\sigma_{\rho}^{1m}\right)^{2} & 0 & \cdots & 0\\
0 & \left(\sigma_{\rho}^{2m}\right)^{2} & \cdots & 0\\
\cdots & \cdots & \cdots & \cdots\\
0 & 0 & \cdots & \left(\sigma_{\rho}^{km}\right)^{2}
\end{array}\right]
\end{align}
where $\left(\sigma_{\rho}^{km}\right)^{2}=\textrm{cov}\left(\varepsilon_{\rho}^{km},\varepsilon_{\rho}^{km}\right)$
represents the ranging error of the $km^{th}$ positioning signal.
Then, the covariance of the positioning error is
\begin{align}
\left(\sigma_{X}^{m}\right)^{2} & =\textrm{cov}\left(\varepsilon_{X}^{m},\varepsilon_{X}^{m}\right)\nonumber \\
 & =H^{m}\left(\sigma_{\rho}^{m}\right)^{2}\left(H^{m}\right)^{T}\label{eq:matrix}
\end{align}
The diagonal elements represent the positioning accuracy of each direction.
Then, the horizontal positioning accuracy can be expressed as
\begin{equation}
\varPsi^{m}=\sqrt{\sum_{k\in\mathcal{K}}\left\{ \left[\sum_{i=1}^{2}\left(\hbar_{ik}^{m}\right)^{2}\right]\left(\sigma_{\rho}^{km}\right)^{2}\right\} }
\end{equation}
where $\hbar_{ik}^{m}$s ($i\in\left\{ 1,2,3\right\} $) represent
the elements of $H^{m}$. Note $\lambda^{km}=\sqrt{\sum_{i=1}^{2}\left(\hbar_{ik}^{m}\right)^{2}}$
as the geometric-dilution, then we have (\ref{eq:1-2}).

\subsection{Derivation of $\left(\sigma_{\rho}^{km}\right)^{2}$\label{subsec:Derivation-of}}

Note
\begin{equation}
A_{0}^{m}=\intop_{B_{0}-B_{fe}/2}^{B_{0}+B_{fe}/2}fG_{p}^{m}\left(f+m\Delta f_{p}\right)\textrm{sin}\left(\pi fDT_{p}\right)df\label{eq:48}
\end{equation}
\begin{equation}
A_{1}^{m}=\intop_{B_{0}-B_{fe}/2}^{B_{0}+B_{fe}/2}N_{0}G_{p}^{m}\left(f+m\Delta f_{p}\right)\textrm{sin}^{2}\left(\pi fDT_{p}\right)df
\end{equation}
\begin{equation}
A_{2}^{m}=\intop_{B_{0}-B_{fe}/2}^{B_{0}+B_{fe}/2}G_{s}^{m}\left(f+m\Delta f_{p}\right)G_{p}^{m}\left(f+m\Delta f_{p}\right)\textrm{sin}^{2}\left(\pi fDT_{p}\right)df
\end{equation}
\begin{equation}
A_{3}^{m}=\intop_{B_{0}-B_{fe}/2}^{B_{0}+B_{fe}/2}G_{q}^{km}\left(f+m\Delta f_{p}\right)G_{p}^{m}\left(f+m\Delta f_{p}\right)\textrm{sin}^{2}\left(\pi fDT_{p}\right)df\label{eq:51}
\end{equation}
\textcolor{black}{Then, (\ref{eq:3}) can be written as
\begin{equation}
\left(\sigma_{\rho}^{km}\right)^{2}=\frac{a\left(A_{1}^{m}+A_{2}^{m}+A_{3}^{m}\right)}{\left(2\pi\right)^{2}\left|h_{p}^{km}\right|^{2}P_{p}^{km}\left(A_{0}^{m}\right)^{2}}\label{eq:52}
\end{equation}
Notice that there are mu}ltiple P-Users, i.e. the bandwidth of the
positioning signal for one P-User is much smaller than the total bandwidth
$B$. Moreover, the front-end bandwidth is larger than $B$ a\textcolor{black}{s
well. So we have $B_{fe}\gg\nicefrac{2}{T_{p}}$. Consequently, a
DLL's narrow early-late spacing $D$ can be applied}\footnote{\textcolor{black}{If $B_{fe}$ is not large enough, the DLL correlation
peak will be flattened which will deteriorate the performance of the
phase discriminator.}}\textcolor{black}{. }When $D\rightarrow0$, $\textrm{sin}\left(\pi fDT_{p}\right)$
in (\ref{eq:48})-(\ref{eq:51}) can be replaced by Taylor expansion
around 0.\textcolor{black}{{} Then, by taking (\ref{eq:5-1}), (\ref{eq:5})-(\ref{eq:8})
into} (\ref{eq:48})-(\ref{eq:51})\textcolor{black}{{} and rearranging
items, we have}
\begin{align}
A_{0}^{m} & =\pi DT_{p}^{2}\intop_{-B_{fe}/2}^{B_{fe}/2}f^{2}\textrm{sinc}^{2}\left(fT_{p}\right)df=\frac{1}{2\pi}DB_{fe}\label{eq:53}
\end{align}
\begin{align}
A_{1}^{m} & =\pi DT_{p}N_{0}A_{0}^{m}
\end{align}
\begin{align}
A_{2}^{m} & =D^{2}T_{p}\sum_{k'\in\mathcal{K}}\sum_{n\in\mathcal{N}}\left|h_{c}^{m\leftarrow k'n}\right|^{2}P_{c}T_{c}\intop_{-B_{fe}/2}^{B_{fe}/2}\textrm{sinc}^{2}\left[\left(f+m\Delta f_{p}-n\Delta f_{c}\right)T_{c}\right]\textrm{sin}^{2}\left(fT_{p}\right)df\nonumber \\
 & \overset{G\gg1}{\approx}D^{2}T_{p}T_{c}P_{c}\sum_{k'\in\mathcal{K}}\sum_{n\in\mathcal{N}}\left|h_{c}^{m\leftarrow k'n}\right|^{2}\textrm{sin}^{2}\left[\pi\left(m-\frac{n}{G}\right)\right]\nonumber \\
 & \thinspace\thinspace\thinspace\thinspace\thinspace\thinspace\thinspace\thinspace\thinspace\thinspace\thinspace\thinspace\thinspace\thinspace\thinspace\thinspace\thinspace\thinspace\thinspace\thinspace\thinspace\thinspace\times\intop_{\left(Gm-n-1\right)\Delta f_{c}}^{\left(Gm-n+1\right)\Delta f_{c}}\textrm{sinc}^{2}\left[\left(f+m\Delta f_{p}-n\Delta f_{c}\right)T_{c}\right]df\nonumber \\
 & \approx D^{2}T_{p}P_{c}\sum_{k'\in\mathcal{K}}\sum_{n\in\mathcal{N}}\left|h_{c}^{m\leftarrow k'n}\right|^{2}\textrm{sin}^{2}\left(\frac{n}{G}\pi\right)
\end{align}
\begin{equation}
A_{3}^{m}=\pi^{2}D^{2}T_{p}^{4}\sum_{k'\in\mathcal{K}^{k}}\left|h_{p}^{k'm}\right|^{2}P_{p}^{k'm}\underset{\bar{A}_{3}}{\underbrace{\intop_{-B_{fe}/2}^{B_{fe}/2}f^{2}\textrm{sinc}^{4}\left(fT_{p}\right)df}}
\end{equation}
where
\begin{align}
\bar{A}_{3} & \overset{B_{fe}\gg2/T_{p}}{\approx}\intop_{-\infty}^{\infty}\frac{\textrm{sin}^{4}\left(\pi fT_{p}\right)}{\pi^{4}f^{2}T_{p}^{4}}df\nonumber \\
 & =\frac{1}{4\pi^{4}T_{p}^{4}}\intop_{-\infty}^{\infty}\left[\frac{4\textrm{sin}^{2}\left(\pi fT_{p}\right)}{f^{2}}-\frac{\textrm{sin}^{2}\left(2\pi fT_{p}\right)}{f^{2}}\right]df\nonumber \\
 & =\frac{1}{2\pi^{2}T_{p}^{3}}\label{eq:57}
\end{align}
Taking (\ref{eq:53})-(\ref{eq:57}) back to (\ref{eq:52}) and rearranging
items, we have
\begin{align}
\left(\sigma_{\rho}^{km}\right)^{2} & \approx\frac{aT_{p}}{2}\left[\frac{N_{0}}{B_{fe}\left|h_{p}^{km}\right|^{2}P_{p}^{km}}\right.\nonumber \\
 & \left.\thinspace\thinspace\thinspace\thinspace\thinspace\thinspace\thinspace\thinspace\thinspace\thinspace\thinspace\thinspace\thinspace\thinspace\thinspace\thinspace\thinspace\thinspace\thinspace\thinspace\thinspace+\frac{2P_{c}\sum_{k'\in\mathcal{K}}\sum_{n=1}^{N}\left|h_{c}^{m\leftarrow k'n}\right|^{2}\textrm{sin}^{2}\left(\frac{n}{G}\pi\right)}{B_{fe}^{2}\left|h_{p}^{km}\right|^{2}P_{p}^{km}}+\frac{\sum_{k'\in\mathcal{K}^{k}}\left|h_{p}^{k'm}\right|^{2}P_{p}^{k'm}}{B_{fe}^{2}\left|h_{p}^{km}\right|^{2}P_{p}^{km}}\right]\nonumber \\
 & =\frac{aT_{p}^{2}}{2}\left(\frac{N_{0}}{B_{fe}T_{p}\left|h_{p}^{km}\right|^{2}P_{p}^{km}}+\frac{BGP_{c}\sum_{k'\in\mathcal{K}}\left|h_{c}^{m\leftarrow k'}\right|^{2}}{B_{fe}^{2}\left|h_{p}^{km}\right|^{2}P_{p}^{km}}+\frac{\sum_{k'\in\mathcal{K}^{k}}\left|h_{p}^{k'm}\right|^{2}P_{p}^{k'm}}{B_{fe}^{2}T_{p}\left|h_{p}^{km}\right|^{2}P_{p}^{km}}\right)
\end{align}
\textcolor{black}{where $\left|h_{c}^{m\leftarrow k'}\right|^{2}=\frac{2}{N}\sum_{n\in\mathcal{N}}\left|h_{c}^{m\leftarrow k'n}\right|^{2}\textrm{sin}^{2}\left(\frac{n}{G}\pi\right)$
is defined as the normalized equivalent channel gain of the communication
signal broadcast by gNB $k'$ to the P-User $m$.}

Let's define\textcolor{black}{{} $\left(C/N_{0}\right)^{km}=\left|h_{p}^{km}\right|^{2}P_{p}^{km}/N_{0}$
as the carrier-to-noise ratio of the $km^{\textrm{th}}$ positioning
signal, }$\left(CPR\right)^{km\leftarrow k'}=\frac{2G\left|h_{c}^{m\leftarrow k'}\right|^{2}P_{c}}{\left|h_{p}^{km}\right|^{2}P_{p}^{km}}$
as the \textcolor{black}{equivalent} communication-to-positioning
ratio of communication signal broadcast by gNB $k'$ to positioning
signal $km$, and $\left(PPR\right)^{km\leftarrow k'm}=\frac{\left|h_{p}^{k'm}\right|^{2}P_{p}^{k'm}}{\left|h_{p}^{km}\right|^{2}P_{p}^{km}}$
as the positioning-to-positioning ratio of the $k'm^{\textrm{th}}$
to the $km^{\textrm{th}}$ positioning signal. Then we have (\ref{eq:7-1}).

\subsection{Derivation of $\tilde{P}_{p}^{km}$\label{subsec:Derivation-of-1}}

The KKT conditions of OP2 can be written as
\begin{eqnarray}
{\displaystyle \sum_{n\in\mathbb{N}_{m}}}\tilde{\mu}^{kn}\left(I_{th}^{kn}-I^{kn}\right) & = & 0\\
\tilde{\beta}^{km}\left(|h_{p}^{km}|^{2}P_{p}^{km}-\varrho\Omega|h_{p}^{k_{k}^{\prime}m}|^{2}P_{p}^{k_{k}^{\prime}m}\right) & = & 0\\
\frac{\partial\tilde{\mathcal{L}}\left(\left\{ P_{p}^{km}\right\} ,\tilde{\mu}^{kn},\tilde{\beta}^{km}\right)}{\partial P_{p}^{km}} & = & 0\label{eq:Piandao}
\end{eqnarray}
\textcolor{black}{It is obvious that the optimal solution $\tilde{P}_{p}^{km}$
satisfies (\ref{eq:Piandao}). }Thus, take (\ref{eq:L2}) into \textcolor{black}{(\ref{eq:Piandao})},
we have
\begin{eqnarray}
\frac{\partial\tilde{\mathcal{L}}}{\partial P_{p}^{km}} & = & \dfrac{-\frac{1}{M}\partial\left(\lambda^{km}\sigma_{\rho}^{km}\right)^{2}-\nu^{k}P_{p}^{km}}{\partial P_{p}^{km}}+\dfrac{\partial{\displaystyle \sum_{n\in\mathbb{N}_{m}}}\tilde{\mu}^{kn}\left(I_{th}^{kn}-I^{kn}\right)}{\partial P_{p}^{km}}\nonumber \\
 &  & +\dfrac{\partial\left\{ \tilde{\beta}^{km}\left(|h_{p}^{km}|^{2}P_{p}^{km}-\varrho\Omega|h_{p}^{k_{k}^{\prime}m}|^{2}P_{p}^{k_{k}^{\prime}m}\right)\right\} }{\partial P_{p}^{km}}\nonumber \\
 & = & -\frac{1}{M}\left(\dfrac{\lambda^{km}\widetilde{\sigma}_{\rho}^{km}}{P_{p}^{km}}\right)^{2}-\nu^{k}-{\displaystyle \sum_{n\in\mathbb{N}_{m}}}\tilde{\mu}^{kn}\underset{J^{kn\leftarrow m}}{\underbrace{\frac{\partial I^{kn}}{\partial P_{p}^{km}}}}+\tilde{\beta}^{km}|h_{p}^{km}|^{2}\label{eq:op2}
\end{eqnarray}
By taking (\ref{eq:2}) into (\ref{eq:op2}), we have
\begin{equation}
J^{kn\leftarrow m}=\sum_{k'\in\mathcal{K}}\left|h_{p}^{kn\leftarrow k'm}\right|^{2}T_{p}\textrm{sinc}^{2}\left(m-\frac{n}{G}\right)\label{eq:56}
\end{equation}
Then, by setting (\ref{eq:op2}) to 0, we can obtain the optimal power
allocation solution as (\ref{eq:P_end}) shows.

\subsection{Subgradient Method of the Lagrange Dual Function\label{subsec:Subgradient-method-of}}

For a set of dual variable $\left\{ \tilde{\mu},\tilde{\nu},\tilde{\beta}\right\} $,
it is known that if $g^{k}\left(\tilde{\mu}^{kn}\right)\geq g^{k}\left(\mu^{kn}\right)+s\left(\tilde{\mu}^{kn}-\mu^{kn}\right)$
holds for any feasible $\tilde{\mu}^{kn}$, then $s$ must be the
subgradient of $g^{k}\left(\mu^{kn}\right)$ at $\mu^{kn}$. Then
the Lagrange dual function of sub-problem (\ref{eq:Fen_1}) is
\begin{eqnarray}
g^{k}\left(\tilde{\mu},\tilde{\nu},\tilde{\beta}\right) & = & \underset{P_{p}^{km}}{\max}\mathcal{L}\left(\left\{ P_{p}^{km}\right\} ,\tilde{\mu},\tilde{\nu},\tilde{\beta}\right)\nonumber \\
 & = & \underset{P_{p}^{km}}{\max}\left[-\frac{1}{M}\sum_{m\in\mathcal{M}}\lambda^{km}\left(\widetilde{\sigma}_{\rho}^{km}\right)^{2}+\sum_{n\in\mathcal{N}}\tilde{\mu}^{kn}\left(I_{th}^{kn}-I^{kn}\right)\right.\nonumber \\
 &  & \left.+\tilde{\nu}^{k}\left(P_{th}^{k}-{\displaystyle \sum_{m\in\mathcal{M}}P_{p}^{km}}\right)+\sum_{m\in\mathcal{M}}\tilde{\beta}^{km}\left(|h_{p}^{km}|^{2}P_{p}^{km}-\varrho\Omega|h_{p}^{k_{k}^{\prime}m}|^{2}P_{p}^{k_{k}^{\prime}m}\right)\right]\nonumber \\
 & \geq & -\frac{1}{M}\sum_{m\in\mathcal{M}}\lambda^{km}\left(\widetilde{\sigma}_{\rho}^{km}\right)^{2}+\sum_{n\in\mathcal{N}}\tilde{\mu}^{kn}\left(I_{th}^{kn}-I^{kn}\right)\nonumber \\
 &  & +\tilde{\nu}^{k}\left(P_{th}^{k}-{\displaystyle \sum_{m\in\mathcal{M}}P_{p}^{km}}\right)+\sum_{m\in\mathcal{M}}\tilde{\beta}^{km}\left(|h_{p}^{km}|^{2}P_{p}^{km}-\varrho\Omega|h_{p}^{k_{k}^{\prime}m}|^{2}P_{p}^{k_{k}^{\prime}m}\right)\nonumber \\
 & = & g^{k}\left(\mu,\nu,\beta\right)+\left(\tilde{\nu}^{k}-\nu^{k}\right)\left({\displaystyle P_{th}^{k}-{\displaystyle \sum_{m\in\mathcal{M}}P_{p}^{km}}}\right)+\sum_{n\in\mathcal{N}}\left(\tilde{\mu}^{kn}-\mu^{kn}\right)\left(I_{th}^{kn}-I^{kn}\right)\nonumber \\
 &  & +\sum_{m\in\mathcal{M}}\left(\tilde{\beta}^{km}-\beta^{km}\right)\left(|h_{p}^{km}|^{2}P_{p}^{km}-\varrho\Omega|h_{p}^{k_{k}^{\prime}m}|^{2}P_{p}^{k_{k}^{\prime}m}\right)\label{eq:tidu}
\end{eqnarray}
\bibliographystyle{IEEEtran}
\bibliography{Ref}

\begin{thebibliography}{10}
\providecommand{\url}[1]{#1}
\csname url@samestyle\endcsname
\providecommand{\newblock}{\relax}
\providecommand{\bibinfo}[2]{#2}
\providecommand{\BIBentrySTDinterwordspacing}{\spaceskip=0pt\relax}
\providecommand{\BIBentryALTinterwordstretchfactor}{4}
\providecommand{\BIBentryALTinterwordspacing}{\spaceskip=\fontdimen2\font plus
\BIBentryALTinterwordstretchfactor\fontdimen3\font minus
  \fontdimen4\font\relax}
\providecommand{\BIBforeignlanguage}[2]{{%
\expandafter\ifx\csname l@#1\endcsname\relax
\typeout{** WARNING: IEEEtran.bst: No hyphenation pattern has been}%
\typeout{** loaded for the language `#1'. Using the pattern for}%
\typeout{** the default language instead.}%
\else
\language=\csname l@#1\endcsname
\fi
#2}}
\providecommand{\BIBdecl}{\relax}
\BIBdecl

\bibitem{del2018Survey}
J.~A. del Peral-Rosado, R.~Raulefs, J.~A. Lopez-Salcedo, and G.~Seco-Granados,
  ``Survey of cellular mobile radio localization methods: from 1{G} to 5{G},''
  \emph{IEEE Communications Surveys and Tutorials}, vol.~PP, no.~99, pp. 1--1,
  2018.

\bibitem{Shojafar2016Energy}
M.~Shojafar, N.~Cordeschi, and E.~Baccarelli, ``Energy-efficient adaptive
  resource management for real-time vehicular cloud services,'' \emph{IEEE
  Transactions on Cloud Computing}, vol.~PP, no.~99, pp. 1--1, 2016.

\bibitem{Liang2012Automated}
H.~Liang, G.~X. Gao, T.~Walter, and P.~Enge, ``Automated verification of
  potential {GPS} signal-in-space anomalies using ground observation data,'' in
  \emph{Position Location and Navigation Symposium}, 2012.

\bibitem{7279407}
L.~{Deng} and S.~{Ye}, ``Vehicle information system design based on {B}eidou
  navigation satellite system,'' in \emph{IEEE International Conference on
  Information and Automation}, Aug 2015, pp. 867--870.

\bibitem{Yin2018A}
L.~Yin, Q.~Ni, and Z.~Deng, ``A {GNSS/5G} integrated positioning methodology in
  {D2D} communication networks,'' \emph{IEEE Journal on Selected Areas in
  Communications}, vol.~36, no.~2, pp. 351--362, 2018.

\bibitem{Jeon2016An}
J.~Jeon, Y.~Kong, Y.~Nam, and K.~Yim, ``An indoor positioning system using
  {B}luetooth {RSSI} with an accelerometer and a barometer on a smartphone,''
  in \emph{International Conference on Broadband and Wireless Computing,
  Communication and Applications}, 2016, pp. 528--531.

\bibitem{Tomic2014RSS}
S.~Tomic, M.~Beko, and D.~Rui, ``{RSS}-based localization in wireless sensor
  networks using convex relaxation: Noncooperative and cooperative schemes,''
  \emph{IEEE Transactions on Vehicular Technology}, vol.~64, no.~5, pp.
  2037--2050, 2014.

\bibitem{Yin2019Intelligent}
L.~{Yin}, Q.~{Ni}, and Z.~{Deng}, ``Intelligent multisensor cooperative
  localization under cooperative redundancy validation,'' \emph{IEEE
  Transactions on Cybernetics}, pp. 1--13, 2019.

\bibitem{ZouWinIPS}
H.~{Zou}, M.~{Jin}, H.~{Jiang}, L.~{Xie}, and C.~J. {Spanos}, ``Win{IPS}:
  {W}i{F}i-based non-intrusive indoor positioning system with online radio map
  construction and adaptation,'' \emph{IEEE Transactions on Wireless
  Communications}, vol.~16, no.~12, pp. 8118--8130, Dec 2017.

\bibitem{Shuai2016Automatic}
H.~Shuai, Z.~Gong, W.~Meng, L.~Cheng, D.~Zhang, and W.~Tang, ``Automatic
  precision control positioning for wireless sensor network,'' \emph{IEEE
  Sensors Journal}, vol.~16, no.~7, pp. 2140--2150, 2016.

\bibitem{Chen2018Coverage}
J.~{Chen}, X.~{Ge}, and Q.~{Ni}, ``Coverage and handoff analysis of 5{G}
  fractal small cell networks,'' \emph{IEEE Transactions on Wireless
  Communications}, vol.~18, no.~2, pp. 1263--1276, Feb 2019.

\bibitem{Zhou2018Coverage}
Y.~Zhou, V.~W.~S. Wong, and R.~Schober, ``Coverage and rate analysis of
  millimeter wave {NOMA} networks with beam misalignment,'' \emph{IEEE
  Transactions on Wireless Communications}, vol.~17, no.~12, pp. 1--1, 2018.

\bibitem{Vaghefi2016Cooperative}
R.~M. Vaghefi and R.~M. Buehrer, ``Cooperative {UTDOA} positioning in {LTE}
  cellular systems,'' in \emph{Globecom Workshops}, 2016.

\bibitem{Cui2017Real}
X.~Cui, A.~Gulliver, H.~Song, and J.~Li, ``Real-time positioning based on
  millimeter wave device to device communications,'' \emph{IEEE Access},
  vol.~4, no.~99, pp. 5520--5530, 2017.

\bibitem{3GPPTR38855}
3GPPTR38.855v.16.0.0, ``Study on {NR} positioning support,'' 2019.

\bibitem{Schloemann2015A}
J.~Schloemann, H.~S. Dhillon, and R.~M. Buehrer, ``A tractable analysis of the
  improvement in unique localizability through collaboration,'' \emph{IEEE
  Transactions on Wireless Communications}, vol.~15, no.~6, pp. 3934--3948,
  2015.

\bibitem{Deng2013Situation}
Z.~Deng, Y.~U. Yanpei, Y.~Xie, and N.~Wan, ``Situation and development tendency
  of indoor positioning,'' \emph{China Communications}, vol.~10, no.~3, pp.
  42--55, 2013.

\bibitem{Manna2016Performance}
T.~Manna and A.~Kole, ``Performance analysis of secure {DSSS} multiuser
  detection under near-far environment,'' in \emph{International Conference on
  Intelligent Control Power and Instrumentation}, 2016.

\bibitem{Schloemann2015AT}
J.~Schloemann, H.~S. Dhillon, and R.~M. Buehrer, ``A tractable metric for
  evaluating base station geometries in cellular network localization,''
  \emph{IEEE Wireless Communications Letters}, vol.~5, no.~2, pp. 140--143,
  2015.

\bibitem{Chen2015Enhanced}
C.~H. Chen and K.~T. Feng, ``Enhanced distance and location estimation for
  broadband wireless networks,'' \emph{IEEE Transactions on Mobile Computing},
  vol.~14, no.~11, pp. 1--1, 2015.

\bibitem{Yu2018Link}
W.~{Yu}, L.~{Musavian}, and Q.~{Ni}, ``Link-layer capacity of {NOMA} under
  statistical delay {Q}o{S} guarantees,'' \emph{IEEE Transactions on
  Communications}, vol.~66, no.~10, pp. 4907--4922, Oct 2018.

\bibitem{Liu2017Nonorthogonal}
Y.~Liu, Z.~Qin, M.~Elkashlan, Z.~Ding, A.~Nallanathan, and L.~Hanzo,
  ``Nonorthogonal multiple access for 5{G} and beyond,'' \emph{Proceedings of
  the IEEE}, vol. 105, no.~12, pp. 2347--2381, 2017.

\bibitem{Yin2019Anovel}
L.~{Yin}, J.~{Cao}, K.~{Lin}, Z.~{Deng}, and Q.~{Ni}, ``A novel
  positioning-communication integrated signal in wireless communication
  systems,'' \emph{IEEE Wireless Communications Letters}, pp. 1--1, 2019.

\bibitem{Karmokar2015Energy}
A.~K. Karmokar, M.~Naeem, and A.~Anpalagan, ``Energy-efficient subcarrier power
  allocation for cognitive radio networks using statistical interference
  model,'' in \emph{IEEE International Symposium on Personal, Indoor, and
  Mobile Radio Communications}, 2015.

\bibitem{6758420}
N.~{Forouzan} and S.~A. {Ghorashi}, ``New algorithm for joint subchannel and
  power allocation in multi-cell {OFDMA}-based cognitive radio networks,''
  \emph{IET Communications}, vol.~8, no.~4, pp. 508--515, March 2014.

\bibitem{Ding2017Joint}
X.~Ding and Q.~Li, ``Joint power control and time allocation for wireless
  powered underlay cognitive radio networks,'' \emph{IEEE Wireless
  Communications Letters}, vol.~PP, no.~99, pp. 1--1, 2017.

\bibitem{Yan2018Toward}
C.~Yan, A.~Bayesteh, Y.~Wu, B.~Ren, S.~Kang, S.~Sun, X.~Qi, Q.~Chen, B.~Yu, and
  Z.~Ding, ``Toward the standardization of non-orthogonal multiple access for
  next generation wireless networks,'' \emph{IEEE Communications Magazine},
  vol.~56, no.~3, pp. 19--27, 2018.

\bibitem{7954630}
J.~{Zhao}, Y.~{Liu}, K.~K. {Chai}, A.~{Nallanathan}, Y.~{Chen}, and Z.~{Han},
  ``Spectrum allocation and power control for non-orthogonal multiple access in
  {H}et{N}ets,'' \emph{IEEE Transactions on Wireless Communications}, vol.~16,
  no.~9, pp. 5825--5837, Sep. 2017.

\bibitem{Zhong2009Geometric}
E.~J. Zhong and T.~Z. Huang, ``Geometric dilution of precision in navigation
  computation,'' in \emph{International Conference on Machine Learning and
  Cybernetics}, 2009.

\bibitem{Liu2016Cooperative}
Y.~Liu, Z.~Ding, M.~Elkashlan, and H.~V. Poor, ``Cooperative non-orthogonal
  multiple access with simultaneous wireless information and power transfer,''
  \emph{IEEE Journal on Selected Areas in Communications}, vol.~34, no.~4, pp.
  938--953, 2016.

\bibitem{Betz2009Generalized}
J.~W. Betz and K.~R. Kolodziejski, ``Generalized theory of code tracking with
  an early-late discriminator {P}art {I}: Lower bound and coherent
  processing,'' \emph{IEEE Transactions on Aerospace and Electronic Systems},
  vol.~45, no.~4, pp. 1538--1556, 2009.

\bibitem{Boyd2006Convex}
Boyd, Vandenberghe, and Faybusovich, \emph{Convex Optimization}, 2004.

\bibitem{Zhang2012Optimal}
R.~{Zhang}, ``Optimal power control over fading cognitive radio channel by
  exploiting primary user {CSI},'' in \emph{IEEE Global Communications
  Conference (GLOBECOM)}, Nov 2008, pp. 1--5.

\bibitem{6775304}
L.~{Li} and C.~{Xu}, ``On ergodic sum capacity of fading channels in
  {OFDMA}-based cognitive radio networks,'' \emph{IEEE Transactions on
  Vehicular Technology}, vol.~63, no.~9, pp. 4334--4343, Nov 2014.

\bibitem{Peral2012Joint}
J.~A. {del Peral-Rosado}, J.~A. {Lopez-Salcedo}, G.~{Seco-Granados},
  F.~{Zanier}, and M.~{Crisci}, ``Joint channel and time delay estimation for
  {LTE} positioning reference signals,'' in \emph{ESA Workshop on Satellite
  Navigation Technologies European Workshop on GNSS Signals and Signal
  Processing}, Dec 2012, pp. 1--8.

\bibitem{Lu2016Quality}
Y.~Lu, Z.~Deng, Z.~Di, and E.~Hu, ``Quality assessment method of {GNSS} signals
  base on multivariate dilution of precision,'' in \emph{European Navigation
  Conference}, 2016.

\end{thebibliography}

\end{document}